\documentclass[pra,floatfix,twocolumn,amsfonts,amsmath,amssymb,nofootinbib,reprint,superscriptaddress]{revtex4-2}
\pdfoutput=1
\usepackage{calligra}
 
\usepackage{graphicx}
\usepackage{dcolumn}
\usepackage{bm}
\usepackage[utf8]{inputenc}
\usepackage[makeroom]{cancel}
\usepackage[colorlinks]{hyperref}
\usepackage[normalem]{ulem}
\usepackage{comment}
\usepackage{xcolor}
\usepackage{amsthm}
\usepackage{mathtools}
\usepackage{bbm}
\usepackage{bm}
\usepackage{graphicx}
\usepackage{enumerate}
\usepackage{tikz}
\usetikzlibrary{arrows}
\usepackage{soul}
\usepackage{float} 

\newtheorem{theorem}{Theorem}
\newtheorem{definition}{Definition}
\newtheorem{proposition}{Proposition}

\newtheorem{corollary}{Corollary}

\newtheorem{lemma}{Lemma}
\newtheorem{obs}{Observation}

\def\ketbra#1#2{\mathinner{|{#1}\rangle\!\langle{#2}|}}

\def\a{\bm{a}}
\def\x{\bm{x}}
\def\w{\bm{w}}
\def\BraVert{\egroup\,|@vertical\,\bgroup}

\DeclareMathOperator{\Tr}{Tr}

\renewcommand\L{\mathcal{L}}

\newcommand{\M}{\mathcal{M}}
\newcommand{\HS}{\mathcal{H}}
\newcommand{\N}{\mathcal{N}}
\newcommand{\A}{\mathcal{A}}

\newcommand{\X}{\mathcal{X}}

\newcommand{\id}{\mathbbm{1}}

\definecolor{cool_green}{rgb}{0.0, 0.5, 0.0}

\begin{document}

\preprint{APS/123-QED}

\title{
Bell scenarios encode classical non-causality
}

\author{{Andreas Leitherer}}
\email{andreas.leitherer@gmail.com}
\affiliation{ICFO-Institut de Ciencies Fotoniques, The Barcelona Institute of Science and Technology,\\ 08860 Castelldefels, Barcelona, Spain}

\author{{Kyrylo Simonov}}
\affiliation{Fakult\"at f\"ur Mathematik, Universit\"at Wien, Oskar-Morgenstern-Platz 1, 1090 Vienna, Austria}

\author{{Ravi Kunjwal}}
\affiliation{Aix-Marseille University, CNRS, LIS, Marseille, France}

\author{Antonio Acín}
\affiliation{ICFO-Institut de Ciencies Fotoniques, The Barcelona Institute of Science and Technology,\\ 08860 Castelldefels, Barcelona, Spain}
\affiliation{ICREA, Passeig Lluis Companys 23, 08010 Barcelona, Spain}

\author{Kuntal Sengupta}
\affiliation{University Grenoble Alpes, Inria, 38000 Grenoble, France}

\author{Hippolyte Dourdent}
 \email{hippolyte.dourdent@uni-muenster.de}
\affiliation{ICFO-Institut de Ciencies Fotoniques, The Barcelona Institute of Science and Technology,\\ 08860 Castelldefels, Barcelona, Spain}
\affiliation{Department for Quantum Technology, Universität Münster,
Heisenbergstraße 11, 48149 Münster, Germany}

\date{\today}

\begin{abstract} 
  Bell and contextuality scenarios are unified within the Acín-Fritz-Leverrier-Sainz hypergraph formalism, in
which measurements are hyperedges. Multipartite measurements are obtained from product
rules that combine independent contextuality scenarios into a single hypergraph. Several such rules
exist: they are structurally distinct, yet yield the same classical, quantum, and boxworld
correlations. 
  Here, we show that this difference is not incidental, unveiling an unexpected connection. The measurements of composite Bell scenarios are exactly the process
functions, the deterministic classical communication protocols that may be non-causal, i.e., lack a definite
causal order without generating paradoxes. Validity of a hyperedge and logical consistency
of a process function are thus one and the same condition. This allows us to characterize process functions by normalization and no-signaling alone, providing transformations that generate non-causal process functions from no-communication ones. The product rules, in turn, order process functions by causal structure, from no-communication to non-causality, which yields a graphical criterion for the latter: a process function is non-causal precisely when its hyperedge is generated by the disjunctive but not the maximal product.
We thereby obtain the first joint hypergraph representation of (non-causal) process functions  and (nonlocal-without-entanglement) measurements, recasting classical non-causality as a non-standard way of composing black boxes.
\end{abstract}

\maketitle

\tableofcontents

\medskip

\section{Introduction}

Quantum theory allows for statistics that admit no classical explanation. Such non-classicality famously manifests itself as \textit{Bell nonlocality} \cite{bell1964einstein, bell1966problem, brunner2014bell} and \textit{Kochen–Specker (KS) contextuality} \cite{ks67, budroni2022kochen}. Bell nonlocality is certified when the correlations of spatially separated parties, though respecting \textit{no-signaling}, cannot be reproduced by any local causal model. The KS theorem is a closely related but originally algebraic result, establishing the impossibility of constructing a deterministic assignment of outcomes to quantum measurements that is compatible with their algebraic properties. 
 These two results are brought together within various frameworks, including ontological \cite{spekkens2005contextuality}, graph-theoretic \cite{cabello2014graph}, and sheaf-theoretic \cite{abramsky2011sheaf} ones.
The latter two are unified in the \textit{Acín-Fritz-Leverrier-Sainz (AFLS) formalism} \cite{acin2015combinatorial} that provides a theory-independent, hypergraph-based approach to contextuality. 

The AFLS framework reformulates contextuality operationally:  collections of measurements are described by hypergraphs, in which vertices represent \textit{events} (measurement outcomes) and hyperedges represent measurements. 
Such a scenario is also called a \textit{test space} \cite{wilce2009test, coecke2000operational}. Contextuality is not a property of the scenario itself, but of the empirical statistics defined on it: the latter are called contextual if they cannot be written as a convex combination of deterministic $\{0,1\}$-valued assignments to events, even though the \textit{no-disturbance principle} \cite{cabello10,ramanathan_12} 
is satisfied, i.e. the statistics of a measurement does not depend on which compatible measurements are performed alongside it. 
In this language, the no-signaling principle and Bell nonlocality are the special cases of no-disturbance and Kochen-Specker contextuality, respectively: the context is fixed by the choices of spatially separated parties, i.e,  multipartite Bell scenarios are built by composing local ones.  

This raises a natural question: given several independent contextuality scenarios, how should one compose them into a single hypergraph representing their joint scenario? The composition is non-trivial and should require that probabilistic models on the composite scenario respect the no-signaling principle. For two scenarios, Foulis and Randall \cite{foulis1981empirical} identified the unique composition rule with this property. For more than two scenarios, however, several rules satisfy no-signaling \cite{acin2015combinatorial, sainz2018multipartite}, forming a family of \textit{Foulis–Randall (FR) products}. Despite their formal differences, these products are observationally equivalent at the level of classical, quantum, and boxworld~\cite{popescu1994quantum, barrett2007information}  correlations, i.e., the respective sets of correlations do not change under the composition rules~\cite{sainz2018multipartite}. The one known exception concerns \textit{almost-quantum correlations} \cite{navascues2015almost}, which coincide across the   \textit{common}, \textit{maximal}, and \textit{disjunctive} FR products but differ for the \textit{minimal} one \cite{sainz2018multipartite}. 
The minimal, common, and maximal products admit an operational reading: the respective hyperedges correspond to measurement protocols, i.e., to a specific order in which the parties choose measurements and communicate their outcomes to one another.  For the disjunctive FR product, however, some hyperedges cannot be interpreted as standard measurement protocols. 
Whether these products carry any further physical significance has not been addressed so far.

As shown here, the answer comes from an a priori independent research line: the study of \textit{indefinite causal order} \cite{costa2026indefinitequantumcausality}. This phenomenon is investigated within the \textit{process matrix formalism} \cite{oreshkov2012quantum}: each party performs ordinary quantum operations in its local laboratory, while these operations are linked by a global process that need not admit any consistent ordering of the parties. Some  processes even violate \textit{causal inequalities} \cite{oreshkov2012quantum, branciard2016simplest}, a feature known as \textit{non-causality}. More specifically, what matters here are \textit{process functions}~\cite{baumeler2016,baumeler16, baumeler2019reversible, tobar2020reversible, dourdent2025paradox}, the deterministic classical analogues of process matrices.
Process functions model all single-round, deterministic classical processes among the parties that are logically consistent, in the sense of not generating time-travel paradoxes  while preserving each party's freedom of choice \cite{baumeler2020equivalence}.
Strikingly, there exist process functions that are non-causal, that is they remain consistent even if they do not admit any consistent ordering of the parties. Several families of non-causal process functions for $n \geq 3$ parties are known~\cite{baumeler14,baumeler2016,af,baumeler2022unlimited,tobar2020reversible,mejdoub25,dourdent2025paradox}, the archetype being the \textit{Lugano (AF/BW) process} \cite{af,baumeler16}.

\section{Results}
\noindent In this work, we uncover a sharp
operational distinction between the different FR products, rooted in causality: each measurement they generate defines a process function, with each FR product singling out a different class of causal structures, which form a strict hierarchy.
\medskip

After a general presentation of the AFLS formalism (Section~\ref{section:afls_framework}), we begin, for ease of exposition, with the bipartite case, where we establish a one-to-one correspondence between the hyperedges generated by the original FR product (defined in Section~\ref{sec:bp_bell}) and bipartite process functions, i.e. no-communication and one-way-communication processes (Section~\ref{subsec:bipartiteequiv}, Proposition~\ref{prop:eq_hyedges_bp_pfs}). We then turn to the general case. Relying on a recently introduced combinatorial characterization of process functions \cite{dourdent2026makescausalloopconsistent} as complete pairwise exclusive sets of multipartite events, we show that process functions and the hyperedges generated by the disjunctive FR product \cite{sainz2018multipartite} of local unipartite Bell scenarios are one and the same objects (Section~\ref{sec:equ_pf_hedges}, Theorem~\ref{thm:uap_pf}). 

We then explore three  implications of this equivalence (Section~\ref{sec:applications}).

First, it implies that the validity criteria of process functions and hyperedges of the disjunctive FR product coincide (Section~\ref{sec:ns_characterization}, Corollary~\ref{cor:eq_validity_criteria}). In particular, it recasts pairwise exclusive completeness \cite{dourdent2026makescausalloopconsistent} as a characterization of process functions in terms of normalization and  no-signaling conditions (Section~\ref{sec:ns_characterization}, Corollary~\ref{cor:ns_characterization}).  

Second, it sheds new light on the hierarchy of composite contextuality scenarios generated by the different FR products \cite{acin2015combinatorial, sainz2018multipartite} (Section~\ref{section:hierarchy_FR}), which turns out to mirror a hierarchy of causal structures. The \textit{minimal} product corresponds to space-like separated parties sharing a common future; the \textit{common} product additionally includes all fixed causal orders; the \textit{maximal} product further encompasses all causal process functions, including deterministic classical control of causal order; and the \textit{disjunctive} product finally admits non-causal structures. 
This yields  a graphical criterion for non-causality: a process function is non-causal precisely when the corresponding hyperedge is generated by the disjunctive but not the maximal FR product (Corollary~\ref{cor:causal_witness}). Moreover, we relate this hierarchy to that of quantum circuits with quantum control of causal order (QC-QC)~\cite{wechs2021}.

Third, inspired by previous work~\cite{eftaxias2023multisystem,sainz2018multipartite}, we indicate how the
normalization and no-signaling conditions characterizing process functions  yield \textit{no-signaling substitution rules} which allow to generate non-causal process functions from no-communication ones (Section~\ref{sec:nc_from_ns_derivation}).
 As an example, starting from a minimal hyperedge associated with a no-communication process function, we derive the hyperedge associated with the Lugano process.

\medskip

Taken together, our  results provide the first joint hypergraph representation of classical non-causality and nonlocality without entanglement \textendash\ two
phenomena whose equivalence was established in the case of quantum measurements in~\cite{kunjwal2023nonclassicality, dourdent2025paradox}. They also reveal that classical non-causality can be understood \textendash\ in line with previous work~\cite{eftaxias2023multisystem} \textendash\ as a non-standard way of composing  no-signaling measurements, complementing its usual reading as the classical counterpart of quantum indefinite causal order~\cite{oreshkov2012quantum,costa2026indefinitequantumcausality}.

\section{The AFLS hypergraph framework\label{section:afls_framework}}

Following \cite{acin2015combinatorial}, we first recall the AFLS framework.

\begin{figure*}[ht]
	\begin{center}
	\includegraphics[width=.9\textwidth]{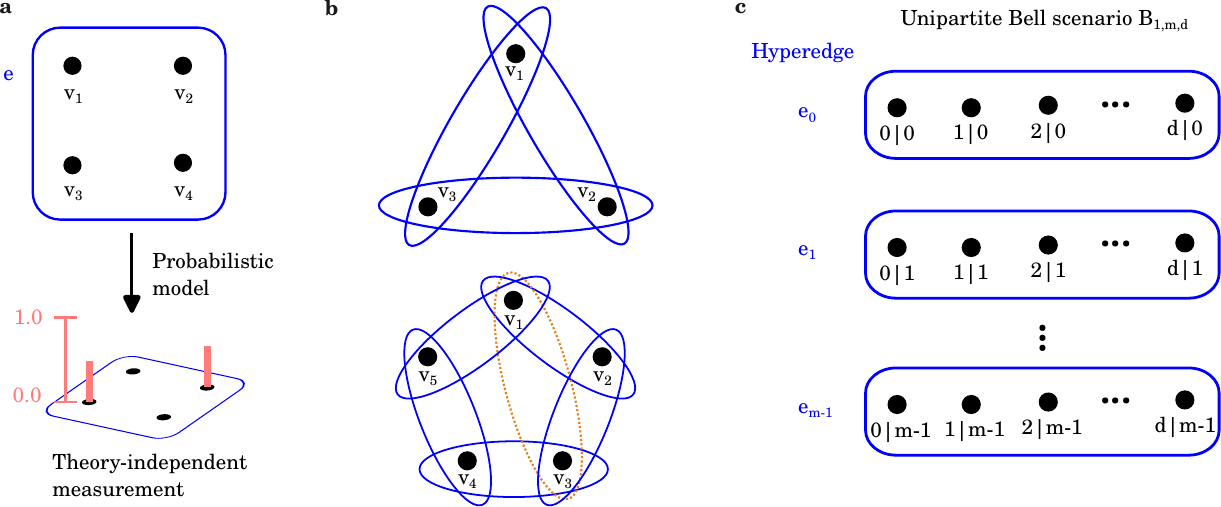}
	\end{center}
	\caption{The AFLS hypergraph framework for a single party. (\textbf{a}) A contextuality scenario is described by a hypergraph, defined by hyperedges that connect a set of vertices. Here, a single hyperedge $e$ (blue box) is shown, which contains four vertices $v_i, i=1,..,4$. $e$ 
    is interpreted as a   measurement 
    and the vertices as its outcomes. A probabilistic model (lower part of (\textbf{a})) assigns a probability to each vertex such that normalization is satisfied, as indicated by the bars over the vertices and the red legend.  This defines a  measurement in a theory-independent way. (\textbf{b}) shows two examples of contextuality scenarios that are contained in the AFLS approach: outcomes are shared across  hyperedges. 
    The pentagon scenario (bottom) also allows for \textit{virtual hyperedges} (orange dashed line), which are implied by the  normalization conditions (cf. Eq.~\eqref{eq:normalization} and Appendix~\ref{appendix:virtual_hyperedges}). Although virtual hyperedges are not required for the main results of this work, they need to be considered for general contextuality scenarios. Additional details are provided in Appendix~\ref{appendix:virtual_hyperedges}.
    (\textbf{c}) A unipartite Bell scenario $B_{1,m,d}$ consists of $m$ disjoint hyperedges  $e_x$, with $x \in \{0,\ldots,m-1$\}, corresponding to $m$ measurement settings, each containing $d$ vertices or rather measurement outcomes.  Each vertex is labeled by the measurement outcome and the associated setting.}
\label{fig:AFLS_introduction}
\end{figure*}

The AFLS framework relies on the notion of hypergraphs.
\begin{definition}[Hypergraph] 
    A \emph{hypergraph} is a pair $H = (V(H), E(H))$, where $V(H)$ is a finite set of vertices, and $E(H) \subseteq 2^{V(H)} \setminus \emptyset$ is a collection of non-empty subsets of $V$, called \emph{hyperedges}.
\label{def:hypergraph}
\end{definition}
Unlike ordinary graphs, whose edges connect exactly two vertices, the hyperedges defining a hypergraph may contain an arbitrary number of vertices in general. Each hyperedge then represents a measurement, i.e., a set of mutually connected vertices corresponding to its mutually exclusive measurement outcomes (cf. Fig.~\ref{fig:AFLS_introduction}a, top). The hyperedges define a notion of orthogonality: two vertices $v_i$ and 
$v_j$ are orthogonal whenever some hyperedge $e$ contains both vertices \textendash that is, whenever they are alternative outcomes of the same measurement. Orthogonal vertices can thus be interpreted as mutually contradictory counterfactual possibilities. 
Different measurements may share outcomes (cf. Fig.~\ref{fig:AFLS_introduction}b): the same outcome can appear in several hyperedges, each corresponding to a different measurement context in which it is obtained. 
A contextuality scenario is then defined as a specific type of hypergraph: 
\begin{definition}[Contextuality scenario] 
    A \emph{contextuality scenario} is a hypergraph $H = (V(H), E(H))$ 
    such that 
    \begin{equation}\label{eq:contextCond}
        \bigcup_{e\in E(H)} e = V(H). 
    \end{equation}
\label{def:contextuality_scenario}
\end{definition}
The condition in Eq.~\eqref{eq:contextCond} ensures that every outcome belongs to at least one measurement. Contextuality scenarios coincide with the notions of \emph{test spaces}~\cite{wilce2009test, coecke2000operational} and related structures in quantum logic~\cite{greechie1971orthomodular}.

Measurement statistics are described by assigning  a probability  to each outcome. 
This is formalized in the following definition. 

\begin{definition}[Probabilistic model]
    A probabilistic model on a contextuality scenario $H$ is a map $p: (V(H),E(H))\rightarrow [0,1]$
    satisfying the following conditions: \begin{enumerate}[(i)]
    \item \emph{No-disturbance principle.} Every outcome is given a unique probability assignment, i.e. for any event $v$ shared by two measurements $e,e'\in E(H),$
        \begin{equation}
        p(v|e) = p(v|e')=p(v)\quad \forall v\in e\cap e', \label{eq:nodisturb}
    \end{equation}
    such that $p: (V(H),E(H))\rightarrow [0,1]$ can be simplified as $p: V(H)\rightarrow [0,1]$.
    \item \emph{Normalization.} Every measurement $e\in E(H)$ is normalized,
    \begin{equation}
        \sum_{v\in e} p(v) = 1\quad \forall e\in E(H). \label{eq:normalization}
    \end{equation}
    \end{enumerate}
    \label{def:probamodel}
\end{definition}
 In (Fig.~\ref{fig:AFLS_introduction}a, bottom) we provide a probabilistic model satisfying the normalization condition \eqref{eq:normalization}. Note that condition \eqref{eq:nodisturb} is implied by the original definitions  provided in \cite{acin2015combinatorial}, while it is not explicitly mentioned.
In a scenario with measurement contexts, i.e. where different measurements share outcomes and thus hyperedges overlap (cf.  Fig.~\ref{fig:AFLS_introduction}b, top), it ensures that each vertex is assigned a unique probability. 
The set of all probabilistic models on $H$ is denoted by $\mathcal{G}(H)$. 

Notably, the AFLS formalism is independent of the theory that defines $\mathcal{G}(H)$: different physical theories simply correspond to different classes of probabilistic models on the same scenario. Classical models, for instance, are convex combinations of deterministic assignments, while quantum models arise from projective measurements on a Hilbert space.  Kochen-Specker contextuality is then naturally represented: 
the Kochen-Specker theorem \cite{ks67} states that some scenarios with overlapping hyperedges admit a quantum probabilistic model but no classical ones \cite{acin2015combinatorial, budroni2022kochen}.
\medskip

In this work, we focus  on \textit{Bell scenarios}:  multipartite no-signaling scenarios where each local party has access to a certain number of independent local measurements. A \textit{unipartite Bell scenario}, denoted by $B_{1,m,d}$, is defined by $m$ measurement settings, each with $d$ outcomes,  with distinct measurements (or rather local operations) sharing no  outcomes. 
Therefore, as a hypergraph, $B_{1,m,d}$ consists of $m$ disjoint (i.e. non-overlapping) hyperedges, each containing $d$ vertices. We label each vertex by the corresponding \textit{unipartite event} $(a|x)$, specifying the outcome $a\in[d]$ obtained upon choosing the measurement setting $x\in[m]$, where $[k] = \{ 0, \ldots , k-1 \}$ (cf. Fig.~\ref{fig:AFLS_introduction}c). 
Note that for the results of this work, it suffices to restrict to the simpler case of each measurement having the same number of outcomes. The more general case of setting-dependent number of outcomes is naturally contained in the AFLS picture, by simply enlarging the number of vertices for each  hyperedge.

\begin{figure}[t!]
	\begin{center}
	\includegraphics[width=.8\columnwidth]{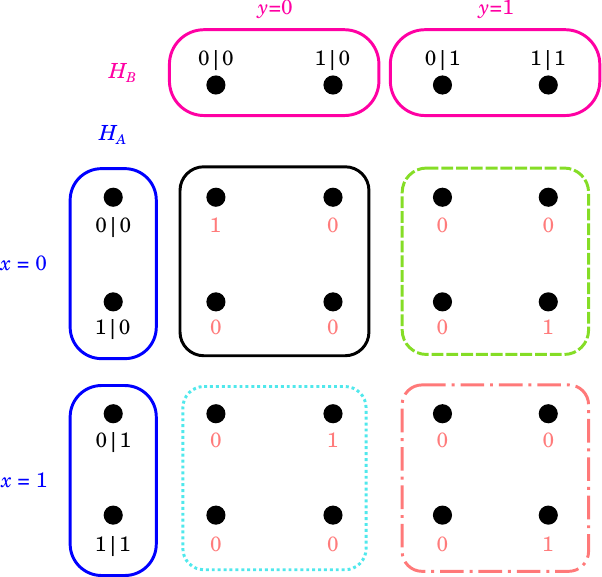}
	\end{center}
	\caption{
    The Cartesian product of two local contextuality scenarios leads to signaling. 
    Two unipartite Bell  scenarios $H_A$ and $H_B$, defined by two measurements settings each having two outcomes, are composed via the Cartesian product $H_A\times H_B=(V_A\times V_B, E(H_A)\times E(H_B))$. One can define a  deterministic probabilistic model compatible with the global hypergraph (the probabilities assigned to the vertices are indicated in light red) such that the marginal outcome probabilities for $A$ depend on the measurement setting of $B$. This is the case, for instance, in the first row for outcome $0$ and setting $0$ for A.    }\label{fig:ambigous_fr_product}
\end{figure}

\begin{figure*}[ht]
	\begin{center}
	\includegraphics[width=.99\textwidth]{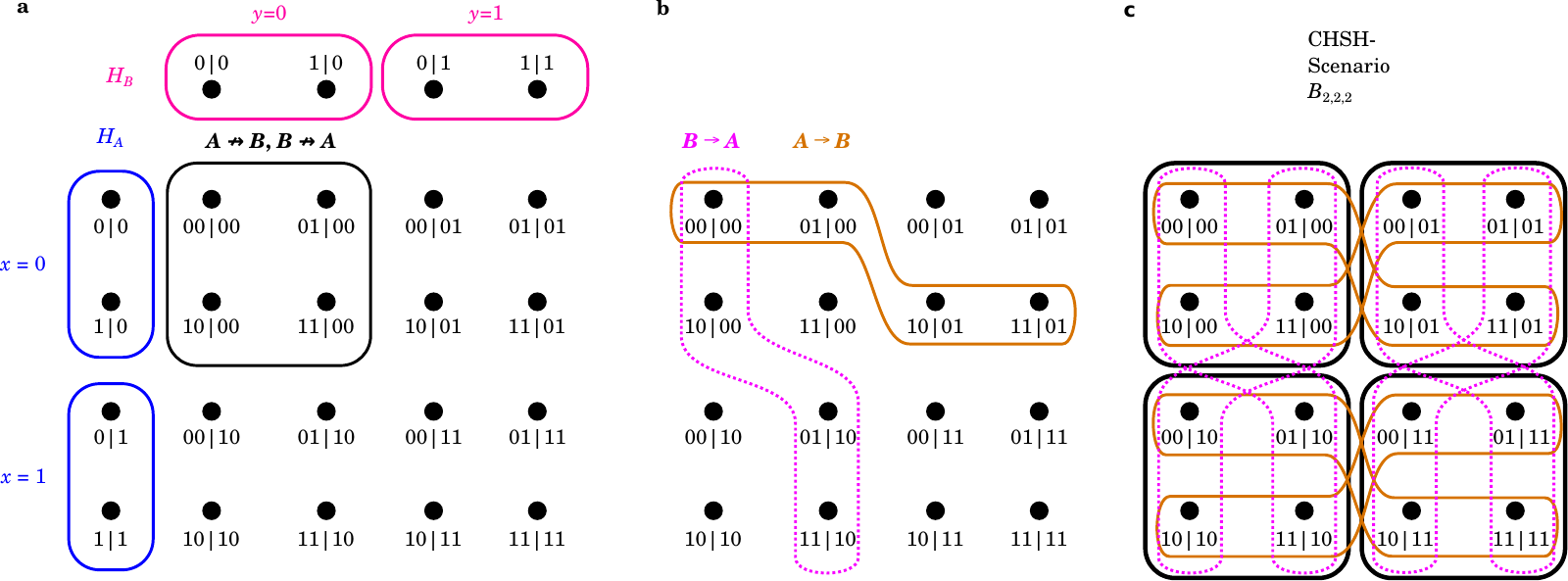}
	\end{center}
	\caption{Recovering the CHSH scenario   from the bipartite Foulis-Randall (FR) product. 
    (\textbf{a}) Two  unipartite Bell scenarios (cf. Fig.~\ref{fig:AFLS_introduction}c) $H_A$ for party $A$ and $H_B$ for party $B$ are depicted as rounded rectangles on the left and top, with the respective local hyperedges (labeled by the measurement setting) colored in blue ($H_A$) and magenta ($H_B$). The composite vertex set is computed via the Cartesian product. A composite, no-communication hyperedge is depicted as a black square (marked by $A\nrightarrow B$, $B\nrightarrow A)$, where both $A$ and $B$ choose the measurement settings zero. 
    (\textbf{b}) shows hyperedges where  
    $A$ signals to $B$ ($A\rightarrow B$, light-brown solid line) and 
    $B$ signals to $A$ ($B\rightarrow A$, violet dashed line).
    (\textbf{c}) Depiction of the complete CHSH scenario, showing   
    all hyperedges that are compatible with normalization and no-signaling, encompassing  no-communication (black) and one-way communication (violet,  light-brown).} 
\label{fig:FR_CHSH}
\end{figure*}

\section{The Foulis-Randall product generates all bipartite process functions}\label{sec:bipartiteequiv}

We first recall the definition and representation of bipartite Bell scenarios within the AFLS framework in Section~\ref{sec:bp_bell}, and then demonstrate the equivalence of their hyperedges with bipartite process functions in Section~\ref{subsec:bipartiteequiv}.

\subsection{Bipartite Bell scenarios \label{sec:bp_bell}}

To go beyond the unipartite case, one must specify how to compose local contextuality scenarios. Given two scenarios $H_A, H_B$, a composition rule $\otimes$  determines a composite scenario $H_A \otimes H_B$. The vertices of the composite scenario are naturally identified with pairs of local vertices,
\begin{equation}
    V(H_A \otimes H_B) = V(H_A) \times V(H_B)
\end{equation}
while the choice of hyperedges depends on the particular composition rule. A probabilistic model on the composite scenario must satisfy normalization and the no-disturbance principle of Definition~\ref{def:probamodel}. In the composite scenario, the latter becomes the no-signaling principle~\cite{ramanathan_12}: 
marginalizing over one party's measurement must yield a well-defined outcome probability for the other parties, i.e. the marginal statistics of one party do not depend on the other party's measurement choice:
\begin{align}
    \sum_{w\in e} p(v,w) &= \sum_{w\in e^\prime} p(v,w) \quad\forall v \in V(H_A), e, e^\prime \in E(H_B) 
    \\
    \sum_{v\in e} p(v,w) &= \sum_{v\in e^\prime} p(v,w) \quad\forall w \in V(H_B), e, e^\prime \in E(H_A).
    \label{eq:ns_fr}
\end{align} 
For unipartite Bell scenarios $H_A =  B_{1,m_A,d_A}$ and $H_B = B_{1,m_B,d_B}$, whose vertices are labeled by local events $(a|x)$ and $(b|y)$, respectively, a vertex of the composite scenario is labeled by a pair of local events,
\begin{equation}
    \left( (a|x), (b|y) \right) =: (ab|xy), \label{eq:not_local_events}
\end{equation}
i.e., a bipartite event representing the joint occurrence of outcomes $a,b$ for local measurement settings $x,y$.

As shown in Ref.~\cite{acin2015combinatorial}, the naive composition choice, the Cartesian product $H_A \times H_B$, is insufficient \textendash\ it admits probabilistic models that violate no-signaling (Fig.~\ref{fig:ambigous_fr_product}). The reason is that the Cartesian products only encodes  simultaneous, non-communicating measurements (cf. Fig.~\ref{fig:FR_CHSH}a).  Enforcing no-signaling requires enlarging the hyperedge set to include all measurements with one-way no-communication \textendash\ including the ones where one party's setting may depend on the other's outcome \cite{foulis1981empirical}. For instance, Fig.~\ref{fig:FR_CHSH}b 
shows, in brown, one such hyperedge: with $x=0$ fixed, $B$'s setting $y$ depends on $A$'s outcome $a$, giving the events $\{(00|00), (01|00), (10|01), (11|01)\}$. Including this hyperedge rules out the probabilistic model of Fig.~\ref{fig:ambigous_fr_product}, since it assigns that hyperedge a total probability of 2, violating normalization.
This is precisely the content of the Foulis--Randall
product~\cite{foulis1981empirical,acin2015combinatorial}:
\begin{definition}[Foulis--Randall product]\label{def:FRprod}
    The FR product $H_A\prescript{\text{FR}}{}{\otimes}H_B$ of two contextuality
    scenarios $H_A$ and $H_B$ has vertex set
    \begin{equation}
        V(H_A \prescript{\text{FR}}{}{\otimes} H_B) = V(H_A) \times V(H_B)
    \end{equation}
    and edge set
    $E(H_A\prescript{\text{FR}}{}{\otimes}H_B)=E_{B\not\rightarrow A}\cup E_{A\not\rightarrow B}$,
    where 
    \begin{align}
        E_{B\not\rightarrow A} &:= \left\{ \bigcup_{v\in e_A}\{v\} \times f(v) :
            e_A\in E_A,\ f:e_A \rightarrow E_B \right\},\label{eq:hyperedges_bipartite_FR_A_to_B} \\
        E_{A\not\rightarrow B} &:= \left\{ \bigcup_{v\in e_B}f(v) \times \{v\} :
            e_B\in E_B,\ f:e_B \rightarrow E_A \right\}.
        \label{eq:hyperedges_bipartite_FR}
    \end{align}
    \label{def:bipartite_FR}
\end{definition}

\noindent Here $E_{B\not\rightarrow A}$ collects the measurements in which $B$ cannot signal to $A$: the setting of $B$ may depend on the outcome of $A$ (non-constant $f$), but not conversely. This includes both the simultaneous measurements $e_A \times e_B$ (constant $f$, no communication) and those with genuine communication from $A$ to $B$. Symmetrically, $E_{A\not\rightarrow B}$ collects the measurements in which $A$ cannot signal to $B$. The intersection of both edge sets  are the simultaneous measurements $e_A \times e_B$ of the Cartesian product.

Applying the Foulis-Randall composition rule to two unipartite Bell scenarios with two inputs and two outputs yields the CHSH
scenario~\cite{clauser1969proposed} (Fig.~\ref{fig:FR_CHSH}c ; see also Fig.7f of Ref.~\cite{acin2015combinatorial})
\begin{equation}
    B_{2,2,2}=B_{1,2,2}\prescript{\text{FR}}{}{\otimes} B_{1,2,2}.
\end{equation}
The probabilistic models over $H_A\prescript{\text{FR}}{}{\otimes}H_B$ are then
exactly the bipartite no-signaling models, i.e.\ the states of boxworld
(Proposition~3.1.4 in \cite{acin2015combinatorial} and Corollary~3.5 in
\cite{barnum2005influence}), a result that extends to the multipartite setting
(Corollary~3.3.8 in \cite{acin2015combinatorial}).

Note that imposing one-way no-communication  at the level of hypergraphs and imposing causal order via  no-signaling constraints leads to equivalent constraints on  probabilistic models. Consider, for instance, 
the hyperedge set of  Eq.~\eqref{eq:hyperedges_bipartite_FR_A_to_B}, which allows communication from $A$ to $B$. The normalization condition for these hyperedges can be shown to be equivalent to the conditions that  impose no-signaling from B to A. This imposes the causal order $A\prec B$ (i.e., A is in the causal past of B). We refer to \cite{acin2015combinatorial,barnum2005influence}, and section 7.2.2 of \cite{dourdent2022causal} for more details.

\subsection{Equivalence of bipartite process functions and Foulis-Randall hyperedges}\label{subsec:bipartiteequiv}

While Foulis and Randall showed that one-way communication is what allows a bipartite scenario to satisfy no-signaling \textendash\ a fact naturally represented within the AFLS framework \textendash\ the connection between the latter and process functions has not been made before.

\begin{figure}[ht]
	\begin{center}
    \includegraphics[width=.7\columnwidth]{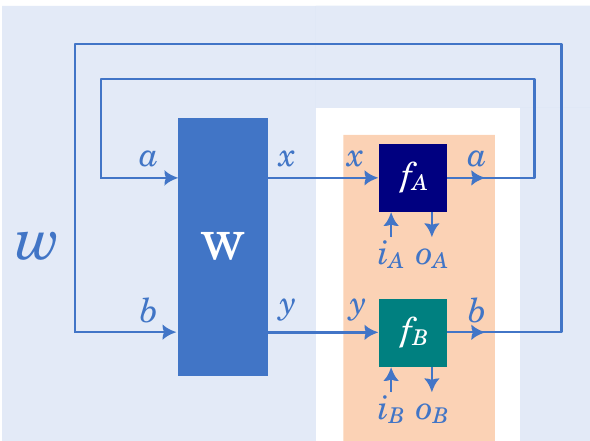}
	\end{center}
	\caption{Sketch of the bipartite process-function communication scenario. Two isolated parties implement arbitrary local interventions $f_A$ and $f_B$. The single-round causal structure between the parties is specified by a process function $w$, represented by a classical channel w together with retro-causal identity channels. The process function provides the incoming variables $x$ and $y$ to parties $A$ and $B$, respectively, while the parties also receive external inputs $i_A$ and $i_B$. For fixed local interventions, the parties generate external outputs $o_A$ and $o_B$ and output variables $a$ and $b$, which are returned to the process function $w$. In this work, following Refs.~\cite{baumeler2019reversible, tobar2020reversible, dourdent2025paradox}, we consider trivial external inputs and outputs, which can thus be neglected.} 
    \label{fig:bipartite_pf}
\end{figure}

Consider a classical communication scenario as depicted in Fig.~\ref{fig:bipartite_pf}, where two parties labeled $A$ and $B$ are embedded in an environment or rather, a shared communication resource.
They receive an input system from the environment, described by  discrete variables  $x\in\mathcal{X}$ for party $A$ and  $y\in\mathcal{Y}$ for party $B$,  and  send back an output variables  $a\in\mathcal{A}$ and  $b\in\mathcal{B}$, respectively. They also receive  external inputs $i_A\in\mathcal{I}_A,i_B\in\mathcal{I}_B$ and produce outcomes $o_A\in\mathcal{O}_A,o_B\in\mathcal{O}_B$  by intervening locally on the input system. Note that 
our notation differs from previous works   on process functions~\cite{kunjwal2023nonclassicality, baumeler2016, tobar2020reversible, baumeler2019reversible, kunjwal2023trading}. 
For each party, the intervention may depend both on the input from the environment ($x$ or $y)$ and the external input ($i_A$ or $i_B$). This is described by the  local stochastic channels $P(a,o_A|x,i_A)$ for party $A$ and  $P(b,o_B|y,i_B)$ for party $B$.
Notably, we restrict to single-round communication, i.e., each party receives, sends and acts on a system only once.
The observed correlations between questions and answers are given by
\begin{equation}
\begin{aligned}
    P(o_A,o_B|i_A,i_B) =
     \sum_{a,b,x,y} & P(a,o_A|x,i_A)P(b,o_B|y,i_B) \\ &\times P(x,y| a,b), \label{eq:correlation_pf_bipartite}
    \end{aligned}
\end{equation}
where $P(x,y| a,b)$ denotes the \textit{classical process} that is carried out by the environment. We restrict our attention to \textit{process functions}: classical shared communication environments that are \textit{logically consistent}, 
i.e.\ never generate paradoxes under arbitrary local operations \cite{baumeler2020equivalence}.
\begin{definition}[Bipartite process function]
    A bipartite process function is a pair of local functions
    $\bm{w}:=(w_A,w_B):\mathcal{A}\times\mathcal{B}\rightarrow\mathcal{X}\times\mathcal{Y}$,
    corresponding to a deterministic classical process
    $P(x,y | a,b)=\delta_{(x,y),\bm{w}(a,b)}$, that is logically consistent:
    for every choice of local operations, $P(a,o_A|x,i_A)$ and $P(b,o_B|y,i_B)$, Eq.~\eqref{eq:correlation_pf_bipartite}
    yields a valid conditional probability distribution $P(o_A,o_B| i_A,i_B)$
    (i.e.\ $P(o_A,o_B| i_A,i_B)\geq 0$ and
    $\sum_{o_A,o_B}P(o_A,o_B| i_A,i_B)=1$ for all $i_A,i_B$).
\end{definition}

The process function scenario depicted in Fig.~\ref{fig:bipartite_pf} also involves deterministic \textit{local interventions}. Following
Refs.~\cite{baumeler2019reversible, tobar2020reversible, dourdent2025paradox}, we take these to have trivial input and output systems $|\mathcal{I}_A|=|\mathcal{O}_A|=1$ and $|\mathcal{I}_B|=|\mathcal{O}_B|=1$, so that they reduce to maps
$f_A:\mathcal{X}\rightarrow\mathcal{A}$ and
$f_B:\mathcal{Y}\rightarrow\mathcal{B}$.

Logical consistency translates into a unique fixed point characterization of process functions \cite{baumeler16,baumeler2019reversible}, which, in the bipartite case, imposes the following constraints:

\begin{theorem}[Characterization of bipartite process functions~\cite{baumeler2019reversible}]\label{th:characbipf}
     A bipartite process function is a pair of local functions
    $\bm{w}:=(w_A,w_B):\mathcal{A}\times\mathcal{B}\rightarrow\mathcal{X}\times\mathcal{Y}$ completely characterized by
    \begin{enumerate}
        \item \emph{No-self-signaling.} The local function producing the input of a party is independent of this party's output:
        \begin{equation}
            \bm{w}(a,b) = \left( w_A(b), w_B(a)\right).
        \end{equation}

        \item \emph{No-two-way-signaling.} At least one of the local function $(w_A,w_B)$ is a constant.
    \end{enumerate} 
\end{theorem}
    
We can now establish the correspondence between bipartite process functions and the FR product of unipartite Bell scenarios. The proof below does not assume any fixed cardinalities for the sets $\mathcal{A}$, $\mathcal{B}$, $\mathcal{X}$, and $\mathcal{Y}$, and therefore applies beyond the Boolean case $B_{2,2,2}$. 
More generally, one could allow the number of outcomes to depend on the measurement setting. However, to establish the correspondence with process functions, it is sufficient to consider the case in which every setting has the same number of outcomes.

\begin{proposition}
Bipartite process functions are in one-to-one correspondence with the hyperedges of the Bell scenario $B_{2,\bm{m},\bm{d}}=B_{1,m_A,d_A}\prescript{\text{FR}}{}{\otimes} B_{1,m_B,d_B}$, where $\bm{m}=(m_A,m_B)$ and $\bm{d}=(d_A,d_B)$. \label{prop:eq_hyedges_bp_pfs}
\end{proposition}

\begin{proof} 
Given a hyperedge of $B_{1,m_A,d_A}\prescript{\mathrm{FR}}{}{\otimes} B_{1,m_B,d_B}$, its vertices $v$ correspond to events $(a,b|x,y)$. Since every output pair $(a,b)$  defining these  events  appears exactly once in the hyperedge, the events define a function
\begin{equation}
\bm{w}=(w_A,w_B):\mathcal{A}\times\mathcal{B}\rightarrow\mathcal{X}\times\mathcal{Y},
\end{equation}
where $\bm{w}(a,b)=(x,y)$ is the unique setting pair associated with the output pair $(a,b)$.

By the definition of the FR product (Def.~\ref{def:FRprod}), every hyperedge is either in $E_{B\not\rightarrow A}$ or in $E_{A\not\rightarrow B}$. Consider first a hyperedge in $E_{B\not\rightarrow A}$. It is of the form
\begin{equation}
    \bigcup_{v\in e_A}\{v\}\times f(v),
\end{equation}
where $e_A$ is an edge of $B_{1,m_A,d_A}$ and $f:e_A\rightarrow E_B$ assigns a local edge of Bob's scenario to each output of Alice. Hence, the corresponding events are $(a,b|x_0,w_B(a))$, 
for some fixed setting $x_0$, implying
\begin{equation}
w_A(a,b)=x_0,\qquad w_B(a,b)=w_B(a).
\end{equation}
Similarly, a hyperedge in $E_{A\not\rightarrow B}$ has the form
\begin{equation}
\bigcup_{v\in e_B}f(v)\times\{v\},
\end{equation}
and gives rise to events $(a,b|w_A(b),y_0)$,
such that
\begin{equation}
w_A(a,b)=w_A(b),\qquad w_B(a,b)=y_0.
\end{equation}
Therefore, every hyperedge defines a pair of local functions satisfying no-self-signaling and no-two-way-signaling. By Theorem~\ref{th:characbipf}, the resulting map $\bm{w}$ is a bipartite process function.

Conversely, let $\bm{w}$ be a bipartite process function. By Theorem~\ref{th:characbipf}, it satisfies $\bm{w}(a,b)=(w_A(b),w_B(a))$, with at least one of the two components being constant. The associated set of events $(a,b|w_A(b),w_B(a))$ therefore has one of the forms
\begin{equation}
(a,b|x_0,y_0),\qquad
(a,b|x_0,w_B(a)),\qquad
(a,b|w_A(b),y_0),
\end{equation}
where $x_0$ and $y_0$ are constants. These are precisely the hyperedges obtained from $E_{B\not\rightarrow A}$ and $E_{A\not\rightarrow B}$ according to Definition~\ref{def:FRprod}. Hence, every bipartite process function defines a hyperedge of $B_{1,m_A,d_A}\prescript{\mathrm{FR}}{}{\otimes} B_{1,m_B,d_B}.$

Bipartite process functions and hyperedges of the FR product are, therefore, equivalent. 
\end{proof}

This proposition establishes a causal interpretation of the FR product, which will be generalized to arbitrary numbers of parties in the following sections. In particular, each hyperedge corresponds to a specific causal order of the parties, as characterized by the associated bipartite process function. Communication hyperedges $A\rightarrow B$ (respectively, $B\rightarrow A$) correspond to process functions with causal order $A\prec B$ (respectively, $B\prec A$), where $A$ lies in the causal past (respectively, $B$ lies in the causal past) of the other party. The no-communication hyperedges correspond to the case in which both causal orders $A\prec B$ and $B\prec A$ are compatible, since neither party's input depends on the other's output.

Moreover,  the necessity of one-way communication for enforcing no-signaling (cf. Section~\ref{sec:bp_bell}) admits a causal reformulation: to impose the bipartite no-signaling principle, one must consider all bipartite process functions, and hence all deterministic classical communications compatible with the no-self-signaling and no-two-way-signaling constraints.

\section{The disjunctive Foulis-Randall product generates all multipartite process functions\label{sec:equ_pf_hedges}}

In the bipartite case, the FR product is uniquely defined. This, however, is not true for the  multipartite case. The reason is that the bipartite FR product is not associative, and hence does not extend uniquely
to more than two parties. Already for three parties it holds that 
\begin{equation}
    H_A \otimes_{\text{FR}} (H_B \otimes_{\text{FR}} H_C) \neq
    (H_A \otimes_{\text{FR}} H_B) \otimes_{\text{FR}} H_C,
    \label{eq:nonassoc_fr}
\end{equation}
since the two orders of composition allow different communication patterns (Proposition~3.3.1
in \cite{acin2015combinatorial}). 

In fact, a whole hierarchy of multipartite products
 arises \cite{acin2015combinatorial, sainz2018multipartite}. This leads to the \textit{minimal, common, maximal,} and the \textit{disjunctive FR product}. On the level of classical, quantum and boxworld models these are all equivalent, while this is not true for all probabilistic model~\cite{sainz2018multipartite}. We discuss the hierarchy in more detail in  Section~\ref{section:hierarchy_FR}.

To generalize Proposition~\ref{prop:eq_hyedges_bp_pfs}, we can employ the nested structure of the hierarchy. 
All four products share the same vertex set, the Cartesian product
$V(H)=\prod_k V(H_k)$; they differ only in their hyperedge
sets, which are nested (Proposition~1 in \cite{sainz2018multipartite}):
\begin{equation}
    \begin{aligned}&E(\prescript{\text{min}}{}{\otimes}_{k}H_k)\subseteq
    E(\prescript{\text{com}}{}{\otimes}_{k}H_k)\subseteq\\&\subseteq
    E(\prescript{\text{max}}{}{\otimes}_{k}H_k)\subseteq
    E(\prescript{\mathrm{disj}}{}{\otimes}_{k}H_k).
    \end{aligned}\label{eq:fr_hierarchy}
\end{equation} 
\noindent If we show that the hyperedges of the largest of these products, the disjunctive FR product, correspond exactly to the set of all process functions, then the hyperedges of every other product in the chain are automatically valid process functions, since they are special cases of those generated by the disjunctive FR product. Hence, we focus on the latter. 
 
 After introducing the multipartite process function framework in Section~\ref{sec:mp_pfs}, and defining the disjunctive FR product in  Section~\ref{sec:disj_frp}, we establish in Section~\ref{sec:eq_dfr_pfs} that its hyperedges precisely characterize the set of multipartite process functions.

\subsection{Multipartite process functions\label{sec:mp_pfs}}

\begin{figure}[ht]
	\begin{center}
	\includegraphics[width=.8\columnwidth]{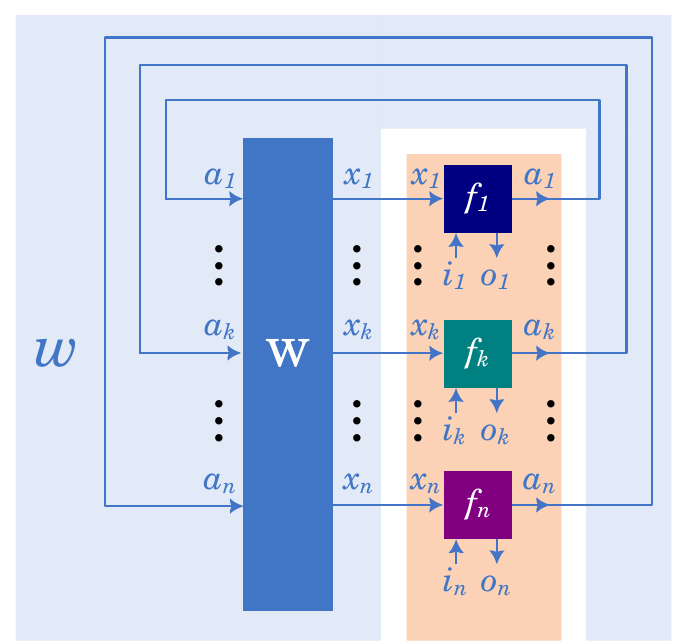}
	\end{center}
	\caption{Sketch of the multipartite process-function communication scenario, generalizing
Fig.~\ref{fig:bipartite_pf} to $n$ parties: each party $k$ performs an arbitrary local
intervention $f_k$, receiving the incoming variable $x_k$ from the process function $w$
and returning the output variable $a_k$. As before, the external inputs $i_k$ and outputs
$o_k$ are taken to be trivial and can thus be neglected~\cite{baumeler2019reversible,
tobar2020reversible, dourdent2025paradox}.}
\label{fig:pf}
\end{figure}

 Consider $n$ parties labeled by $k\in[\![1,n]\!]:=\{1,...,n\}$. 
 We define for each party $k$ the input variables $x_k\in\mathcal{X}_k$, the output variable $a_k\in\mathcal{A}_k$, as well as 
 external inputs $i_k\in\mathcal{I}_k$ and outputs $o_k\in\mathcal{O}_k$ (cf. Fig.~\ref{fig:pf}).
 The local interventions of each party are modeled by  local stochastic channels $P(a_k,o_k|x_k,i_k)$. We emphasize again that we restrict to single-round communications, i.e., each party receives, sends, and acts on a system only once. 
In the following, we will use the shorthand notation 
 $\mathcal{X}:=\bigtimes\limits_k \mathcal{X}_k$, $\mathcal{A}:=\bigtimes\limits_k \mathcal{A}_k$,   $\bm{x}:=(x_1,...,x_n)$, 
$\bm{x}_{\backslash k}:=(x_1,...,x_{k-1},x_{k+1},..,x_n)$,
$\bm{a}:=(a_1,...,a_n)$ etc. 
The observed correlations, as  specified for the bipartite case in Eq.~\ref{eq:correlation_pf_bipartite}, are now
given by 
\begin{equation}
    P(\bm{o}|\bm{i}) = \sum_{\bm{a},\bm{x}} \Bigl(\prod_{k=1}^n P(a_k,o_k| x_k,i_k)\Bigr) P(\bm{x}| \bm{a}), \label{eq:correlation_pf_multipartite}
\end{equation}
where $P(\bm{x}|\bm{a})$ denotes the \emph{classical process} that is carried out by the environment. Multipartite process functions are defined as follows:

\begin{definition}[Multipartite process function]
\label{def:mp_pfs} A process function is 
a set of local 
functions  $\bm{w}:=(w_1, ..., w_n):\mathcal{A}\rightarrow\mathcal{X}$, corresponding to a deterministic classical process   $P(\bm{x}|\bm{a})=\delta_{\bm{x},\bm{w}(\bm{a})}$ that is logically consistent: for every choice of local operations, Eq.~\eqref{eq:correlation_pf_multipartite} yields a valid conditional probability distribution $P(\bm{o}|\bm{i})$ (i.e., $P(\bm{o}|\bm{i}) \geq 0$ and $\sum_{\bm{o}}P(\bm{o}|\bm{i}) = 1$ for all $\bm{i}$).
\end{definition}

The process function scenario depicted in Fig.~\ref{fig:pf} involves deterministic local interventions $\bm{f}:=(f_k)_k$ (where $f_k : \mathcal{X}_k \times \mathcal{I}_k \to \mathcal{A}_k \times \mathcal{O}_k$) such that $P_k(a_k,o_k|x_k,i_k)=\delta_{(a_k,o_k),f_k(x_k,i_k)}$. 
As in the bipartite case, we assume  that 
$|\mathcal{I}_k|=|\mathcal{O}_k|=1$, yielding local interventions $f_k: \mathcal{X}_k\rightarrow \mathcal{A}_k$.

There are several criteria  to validate logical consistency in the multipartite case~\cite{baumeler16,baumeler2019reversible,dourdent2025paradox,dourdent2026makescausalloopconsistent}. Here, we employ a combinatorial, event-based characterization:

\begin{theorem}[Pairwise-exclusive completeness~\cite{dourdent2026makescausalloopconsistent}]\label{thm:characPE}
    A list of events $E\subseteq\mathcal A\times\mathcal X$ is the event list 
    \begin{equation}
\{(\bm{a}|\bm{x})\;|\;\bm a\in\mathcal A, \bm{x}=\bm{w}(\bm{a})\}
\end{equation}     
     of an $n$-partite process function if and only if  \begin{enumerate}[(i)]
    \item $E$ is complete, i.e.,
    \begin{equation}
    |E|=|\mathcal{A}|=\prod_k|\mathcal{A}_k|,
    \end{equation}
    \item Any two distinct events $(\bm{a}|\bm{x})$ and $(\bm{a}'|\bm{x}')$ in $E$ are exclusive,  i.e., there exists a party $k$ such that
    \begin{equation}
    a_k\neq a'_k
    \qquad\text{and}\qquad
    x_k=x'_k . \label{eq:pe}
    \end{equation}
\end{enumerate}
\end{theorem}

Remarkably, for $n\geq3$, there exist process functions that generate correlations incompatible with any well-defined causal order.  
Causal process functions are indeed formalized by the following constraints~\cite{dourdent2025paradox}:
\begin{theorem}
[Characterization of causal process functions~\cite{dourdent2025paradox}]\label{thm:characcausalpf}  
An $n-$partite process function $\bm{w}:\mathcal{A}\to\mathcal{X}$ is \emph{causal} if and only if the following two conditions hold: 
\begin{enumerate}[(i)]
    \item There exists at least one party $k$ such that no other party can signal to it, i.e.~the corresponding input is constant, $ x_{k} = w_{k}(\bm{a}_{\backslash k}) := x_{k}^0 .$

    \item For every such party $k$ and for every $a_{k}\in\mathcal{A}_{k}$, the output-reduced process functions $\bm{w}^{a_{k}}:\mathcal{A}_{\backslash k}\to\mathcal{X}_{\backslash k}$
    defined as
    \begin{align}
        \bm{w}^{a_k}(\bm{a}_{\backslash k}) := ( w_1(\bm{a}_{\backslash 1},a_k), \dots, w_{k-1}(\bm{a}_{\backslash \{k-1\}},a_k),\notag \\
    w_{k+1}(\bm{a}_{\backslash \{k+1\}},a_k), \dots, w_{n}(\bm{a}_{\backslash n},a_k) )\label{eq:red_pf}
    \end{align} 
    are themselves valid $(n-1)-$partite causal process functions.    
\end{enumerate}
\end{theorem}
\noindent A ``non-causal'' process function either has no global past or can be reduced  to a function without a global past. 
The canonical example of a non-causal process function is the so-called \textit{Lugano (AF/BW) process} \cite{af,baumeler2016}, in which each party's input depends non-trivially on the outputs of the other two, so that all three parties lie in each other's past and future:
\begin{align}
    x_1 := a_3(a_2 \oplus 1), \quad x_2 := a_1(a_3 \oplus 1), \quad x_3 := a_2(a_1 \oplus 1),
    \label{eq:lugano}
\end{align}
where $x_1, x_2, x_3$ and $a_1, a_2, a_3$ label the input and output bits for the first, second, and third party, respectively.

\subsection{The disjunctive Foulis-Randall product\label{sec:disj_frp}}

We now recall  the \emph{disjunctive FR product} introduced in Ref.~\cite{sainz2018multipartite}.
 As discussed at the beginning of this section, it contains all the hyperedges of the FR products suggested in the original AFLS framework~\cite{acin2015combinatorial}, and strictly more. Its definition proceeds in three steps.
 First, one needs to introduce a notion of orthogonality for composite events. This turns th set of such events into an orthogonality graph. The \textit{cliques} \textendash\ its sets of mutually connected vertices \textendash\  are precisely the collections of pairwise exclusive events, and thus candidates for valid hyperedges of the composite contextuality scenario. Finally, a clique is considered valid if and only if it is compatible with the normalization and no-signaling conditions.
\medskip

\paragraph{Vertices and orthogonality.}
As discussed after Definition~\ref{def:hypergraph}, the hyperedges of any hypergraph define a notion of orthogonality, which can be formalized as follows:

\begin{definition}[Orthogonality of vertices]
Two distinct vertices $v,w\in V(H)$ are orthogonal in $H$ if and only if they are distinct and belong to a common hyperedge:
\begin{equation}
    v\perp w \quad\Longleftrightarrow\quad v\neq w\ \land\
    \exists\,e\in E(H):\ \{v,w\}\subset e .
\end{equation}
\label{def:orthogonality}
\end{definition}

\begin{figure*}[ht]
	\begin{center}
	\includegraphics[width=.8\textwidth]{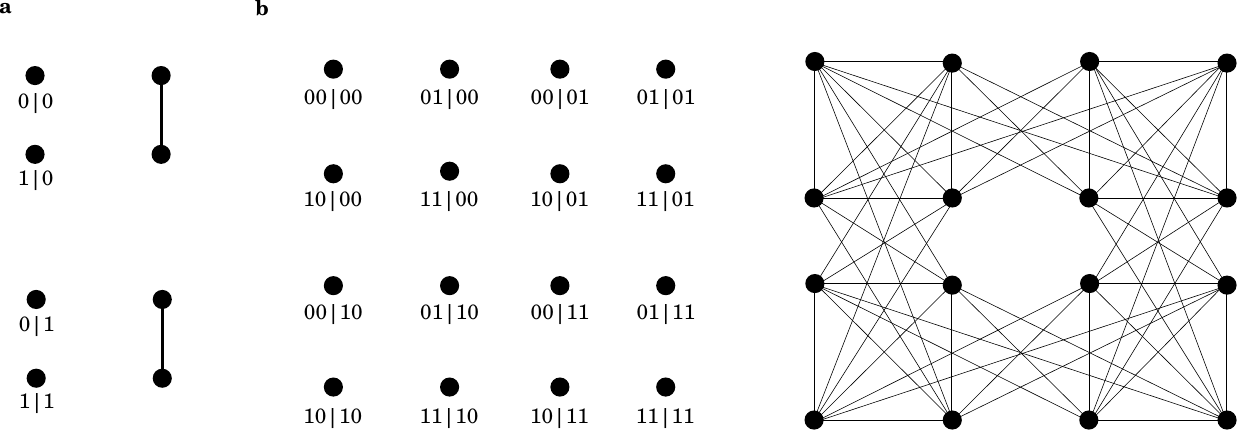}
	\end{center}
	\caption{Scheme of single- and bipartite orthogonality graph (adapted from \cite{acin2015combinatorial}). \textbf{a} shows the vertex set of $B_{1,2,2}$ (left) and its associated orthogonality graph (right). \textbf{b} shows the vertex set of $B_{2,2,2}$ (left) and the associated orthogonality graph (right) which  coincides with the faithful orthogonality graph (cf. Def.~\ref{def:faitfhul_ortho_graph}).}
    \label{fig:Bell_and_Orthograph}
\end{figure*}

Since orthogonality is a binary relation between vertices, it can  be represented by a simple \footnote{The graph does not have more than one edge between any two vertices and no edge starts and ends at the same vertex.} undirected graph (see also Fig.~\ref{fig:Bell_and_Orthograph}).

\begin{definition}[Orthogonality graph]
The orthogonality graph $\mathrm{OrthoGraph}(H)$ associated with a hypergraph $H$ is the simple undirected graph with vertex set $V(H)$ and adjacency relation 
\[
v\sim w
\quad\Longleftrightarrow\quad
v\perp w .
\]
\label{def:orthogonality_graph}
\end{definition}

Multipartite composite scenarios admits an additional and natural notion of orthogonality, that we term \textit{exclusivity}. Noting that the  vertex set of a composite contextuality scenario is given by the Cartesian product $V \left( \bigotimes_{k=1}^{n} H_k \right) = \bigtimes_k V(H_k)$, one may identify two composite vertices to be \textit{exclusive} whenever their local components are orthogonal for at least one party:

 \begin{definition}[Exclusivity]\label{def:pe}
    Two vertices $v = (v_1,...,v_n)$ and $w = (w_1,...w_n)$ of a composite hypergraph   $H=H_1\otimes....\otimes H_n$ are  exclusive if and only if $v$ and $w$ contain local vertices $v_k, w_k$ that are orthogonal:
    \begin{equation}
        v\underset{E}{\perp} w \Leftrightarrow \exists \ k : v_k\perp w_k\ .
    \end{equation}
\end{definition}
\noindent This relation is referred to as \emph{local orthogonality} in Refs.~\cite{acin2015combinatorial,sainz2018multipartite}, following the multipartite principle introduced in Ref.~\cite{fritz2013local}. We use the term  exclusivity to emphasize that we refer only to the underlying combinatorial relation. 

\medskip

\paragraph{The faithful orthogonality graph.}
In general, the orthogonality relations between vertices that are  induced by Def.~\ref{def:orthogonality} and Def.~\ref{def:pe} are not identical. Exclusivity depends only on the local scenarios and not on the particular multipartite composition rule, in contrast to  Def.~\ref{def:orthogonality}. referred to as ``conventional'' orthogonality in \cite{sainz2018multipartite}. Exclusivity is said to be captured by a composition, if the conventional orthogonality relations contain the exclusivity relations. If both notions lead to the same relations, then a composition is said to faithfully capture exclusivity. In particular, a composition satisfies this if and only if it is constructed using a specific graph:

\begin{definition}[Faithful orthogonality graph]
 We define 
 \begin{multline}
    \emph{FaithfulOrthoGraph}(H_1, ..., H_n) \\:= \emph{OrthoGraph}(H_1) * ... * \emph{OrthoGraph}(H_n)
\end{multline}
with the  disjunctive or co-normal graphical product:
\begin{multline}
    (v_1, ..., v_n) \sim (w_1, ..., w_n) \emph{ is edge in } g_1 * ... * g_n \\ \Leftrightarrow\quad \exists k: v_k \sim w_k \emph{ is edge in } g_k.
\end{multline}   
\label{def:faitfhul_ortho_graph}
\end{definition}

By construction, 
two vertices are adjacent in $\mathrm{FaithfulOrthoGraph}(H_1,\ldots,H_n)$ if and only if the corresponding multipartite events are exclusive. Moreover, the disjunctive graph product is associative,
\begin{equation} (
    g_1*g_2)*g_3=g_1*(g_2*g_3),
\end{equation}
i.e., unlike the bipartite FR product (see Eq.~\eqref{eq:nonassoc_fr}), the faithful orthogonality graph is unambiguously defined for any number of parties.

The outcomes of any multipartite measurement must be pairwise exclusive and hence form a \emph{clique} of the faithful orthogonality graph. A clique, however, specifies only a set of mutually exclusive events: pairwise exclusivity alone does not ensure that this set constitutes a complete measurement. Cliques therefore provide candidate hyperedges, and an additional condition is needed to determine which of them define valid multipartite measurements. 

\medskip

\paragraph{Selecting valid hyperedges.}
The candidate hyperedges are filtered by requiring  the corresponding normalization condition to be implied by the operational constraints of the multipartite Bell scenario, namely, normalization and no-signaling. More precisely, a clique $e$ defines a hyperedge  when the condition $\sum_{v\in e} p(v)=1$ is a
linear combination of (i) the \textit{normalization} conditions, one for each
choice of a local measurement made by every party independently (the non-communicating,
Cartesian-product measurements), and (ii) the \textit{no-signaling} equalities,
\begin{equation}
    \sum_{v_k\in e_k} p(\dots,v_k,\dots) =
    \sum_{v_k\in e_k'} p(\dots,v_k,\dots),
    \label{eq:ns_equalities}
\end{equation}
for every party $k$, every pair of local hyperedges $e_k,e_k'\in E(H_k)$, and fixed events of the remaining parties. Equivalently, this condition can be formulated as a rank comparison of the associated $\{0,\pm1\}$-valued matrices~\cite{sainz2018multipartite}; we summarize the corresponding matrix formulation in Appendix~\ref{appendix:matrix_representation}.
\medskip

The above criterion specifies which pairwise-exclusive (in the sense of Def.~\ref{def:orthogonality}) sets of events define valid measurements. 
In particular, it provides the hyperedge structure of the multipartite scenario introduced in Ref.~\cite{sainz2018multipartite}, namely the disjunctive FR product.

\begin{definition}[Disjunctive FR product \cite{sainz2018multipartite}]
    The disjunctive FR product $\prescript{\mathrm{disj}}{}{\otimes}_{k=1}^n H_k$ has
    vertex set $\bigtimes_{k=1}^n V(H_k)$ and, as hyperedges, exactly those cliques
    of $\text{FaithfulOrthoGraph}(H_1,\dots,H_n)$ whose normalization condition
    follows from the normalization and no-signaling constraints.
    \label{def:disjunctive_FR_product_wo_rank}
\end{definition}

\medskip

\begin{obs}\label{obs:exclusivitybell}
    Importantly, for the case of unipartite Bell scenarios $B_{1,m_k,d_k}$, the notion of exclusivity of Definition~\ref{def:orthogonality} coincides with Eq.~\eqref{eq:pe}.
    \begin{proof}
        For $H_k=B_{1,m_k,d_k}$, composite vertices of $H=H_1\otimes ... \otimes H_n$ are represented by events $(\bm{a}|\bm{x}):=(a_1,...,a_n|x_1,...,x_k)$. Two events  $(\bm{a}|\bm{x}), (\bm{a}^\prime|\bm{x}^\prime)$ are exclusive, if and only if there are local events $(a_k|x_k),(a_k^\prime,x_k^\prime)$ that are orthogonal according to Def.~\ref{def:orthogonality}. This enforces that they are in the same local hyperedge, i.e., $x_k=x_k^\prime$, as well as $a_k\neq a_k^\prime$, since otherwise the two local events are identical and thus not orthogonal according to Def.~\ref{def:orthogonality}. 
    \end{proof}
\end{obs}

Notably, the disjunctive FR product of unipartite Bell scenarios contains hyperedges for which no party is
required to measure first. Ref.~\cite{sainz2018multipartite}
provided indeed an explicit example of such a hyperedge, for the case of multipartite Bell scenarios $\prescript{\mathrm{disj}}{}{\otimes}_{k=1}^n B_{1,m_k,d_k}$: 
\begin{align}
    E=\{&
(000|001),\,
(001|101),\,
(010|100),\,
(011|100),\notag\\
&
(100|001),\,
(101|111),\,
(110|101),\,
(111|111)
\}.
\end{align}
We identify these events as the ones constituting the following Lugano-like process\footnote{Via simple local relabelling (for $x_1,x_3$ and $a_3$), one retrieves the standard Lugano process (Eq.~\eqref{eq:lugano}).}
\begin{align}\label{eq:luganob}
    x_1 &= (a_2\oplus 1) (a_3\oplus 1)\oplus 1,\notag\\
    x_2 &= a_3a_1,\notag\\
    x_3 &= (a_1\oplus 1)a_2 \oplus 1
\end{align}

In the next section, we show that the hyperedges of the disjunctive FR product are in one-to-one correspondence with multipartite process functions.

\medskip

\subsection{Equivalence of process functions and disjunctive Foulis-Randall hyperedges \label{sec:eq_dfr_pfs}}

We now come to our main result: the hyperedges of the disjunctive FR product of unipartite Bell scenarios are exactly the process functions. The proof rests on two ingredients. 

On the process-function side, Theorem~\ref{thm:characPE} characterizes them combinatorially: an event list $E\subseteq\mathcal A\times\mathcal X$ is the event list
$\{(\bm a|\bm w(\bm a))\}_{\bm a\in\mathcal A}$ of an $n$-partite process function if and only if it is complete and pairwise exclusive.

On the hypergraph side, we reformulate the validity criterion of the disjunctive FR product on unipartite Bell scenarios \textendash\ which demands that the
normalization of a hyperedge is a linear combination of the normalization and no-signaling 
conditions \textendash\ as the requirement that the
probabilities of its events sum to one for every no-signaling model
(Lemma~\ref{lem:valid_iff_NS_normalized}, Eq.~\eqref{eq:NS_normalized}).

We will then show (Theorem~\ref{thm:uap_pf}) that this validity criterion is precisely pairwise exclusive completeness.

\medskip

Consider
the vertex set of composite events
\begin{equation}\label{eq:vbell}
  V=\bigtimes_{k=1}^n V(B_{1,m_k,d_k})=\bigl\{(\bm a|\bm x)\ :\ \bm a\in\mathcal A,\
  \bm x\in\mathcal X\bigr\},
\end{equation}
and we are interested in which composition of $\bigotimes_{k=1}^n B_{1,m_k,d_k}$, i.e. which partition of $V$ as hyperedges, correspond to the whole set of process functions. The normalization and
no-signaling conditions can therefore be thought, in general, as conditions on assignments $p\in\mathbb R^V$,
independent of any choice of hyperedges. They can be written as
\begin{align}
  \mathrm N_{\bm x}(p)&:=\sum_{\bm a\in\mathcal A}p(\bm a|\bm x)-1 ,
  \label{eq:norm_cond}\\[2pt]
  \mathrm{NS}_i(p)&:=\sum_{a_k}p(a_k\bm a_{\setminus k}| x_k\bm x_{\setminus k})
                    -\sum_{a_k}p(a_k\bm a_{\setminus k}| x_k'\bm x_{\setminus k}) ,
  \label{eq:ns_cond}
\end{align}
where Eq.~\eqref{eq:norm_cond} is indexed by $\bm x\in\mathcal X$ and Eq.~\eqref{eq:ns_cond} by
$i=(k,x_k,x_k',\bm x_{\setminus k},\bm a_{\setminus k})$, expressing that the marginal of
all parties but $k$ does not depend on party $k$'s setting. A \emph{no-signaling model} is
an assignment $p:V\to[0,1]$ with $\mathrm N_{\bm x}(p)=0$ and $\mathrm{NS}_i(p)=0$ for all $\bm x$ and
$i$, i.e. a probabilistic model (Def.~\ref{def:probamodel}) on the vertex set Eq.~\eqref{eq:vbell}.

\begin{lemma}\label{lem:valid_iff_NS_normalized}
A set of events $e\subseteq V$ is a valid hyperedge of
$\prescript{\mathrm{disj}}{}{\otimes}_{k=1}^n B_{1,m_k,d_k}$ if and only if
\begin{equation}\label{eq:NS_normalized}
  \sum_{v\in e}p(v)=1 \qquad\text{for every no-signaling model } p .
\end{equation}
\end{lemma}

\begin{proof}
($\Rightarrow$) Let $e$ be a disjunctive FR hyperedge of
$\prescript{\mathrm{disj}}{}{\otimes}_{k=1}^n B_{1,m_k,d_k}$. By
Def.~\ref{def:disjunctive_FR_product_wo_rank}, its normalization is a linear combination of
the conditions~\eqref{eq:norm_cond}--\eqref{eq:ns_cond}: there exist coefficients
$\lambda_{\bm x},\mu_i\in\mathbb R$ such that
\begin{equation}\label{eq:lincomb}
  \sum_{v\in e}p(v)-1
  =\sum_{\bm x}\lambda_{\bm x}\,\mathrm N_{\bm x}(p)+\sum_i\mu_i\,\mathrm{NS}_i(p),
  \quad\forall p\in\mathbb R^V .
\end{equation}If $p$ is a no-signaling model, by definition, every term on the right-hand side vanishes, which gives
Eq.~\eqref{eq:NS_normalized}.

\medskip\noindent
($\Leftarrow$) Let $e\subseteq V$ satisfy Eq.~\eqref{eq:NS_normalized}. We verify the two
requirements of Def.~\ref{def:disjunctive_FR_product_wo_rank}.

\emph{(i) $e$ is a clique of the orthogonality graph.}
Recall that a \emph{local
deterministic model} is specified by the choice of a single response function
$f_k:\mathcal X_k\to\mathcal A_k$ per party, where each party $k$ outputs $f_k(x_k)$ on setting $x_k$ with certainty, i.e.,  for  a fixed string of measurement settings $\bm{x}$, only one joint outcome $\bm{a}$ appears with certainty:
\begin{equation}\label{eq:local_det_model}
  p_{\bm f}(\bm a|\bm x)=\prod_{k=1}^n\bigl[a_k=f_k(x_k)\bigr]\in\{0,1\} .
\end{equation}
Such models are local, hence no-signaling, so Eq.~\eqref{eq:NS_normalized} applies to them.

Assume, \textit{ad absurdum}, that two distinct events
$(\bm a|\bm x),(\bm a'|\bm x')\in e$ are not exclusive. By
Observation~\ref{obs:exclusivitybell}, exclusivity would require a party $k$ with
$x_k=x_k'$ and $a_k\neq a_k'$; its failure therefore means that for every party $k$ such that $x_k=x_k'$, we have $a_k=a_k'$, and thus, in this case, $f_k(x_k)=f_k(x_k')=a_k=a_k'$. Choosing the responses to the remaining settings
arbitrarily, we obtain a local deterministic model Eq.~\eqref{eq:local_det_model} that assigns probability $1$ to both
events, $p_{\bm f}(\bm a|\bm x)=p_{\bm f}(\bm a'|\bm x')=1$, so that $\sum_{v\in e}p_{\bm f}(v)\ge 2$, contradicting Eq.~\eqref{eq:NS_normalized}. Hence the events
of $e$ are pairwise exclusive.

\emph{(ii) The normalization of $e$ follows from \eqref{eq:norm_cond}--\eqref{eq:ns_cond}.}
These conditions cut out the affine subspace
$\mathsf A=\{p\in\mathbb R^V: \mathrm N_{\bm x}(p)=\mathrm{NS}_i(p)=0\ \forall \bm x,i\}$,
whose elements need not be non-negative; the no-signaling models are precisely those that
are, $\mathcal{NS}=\{p\in\mathsf A: p\ge0\}$. By Eq.~\eqref{eq:NS_normalized}, the affine map
$p\mapsto\sum_{v\in e}p(v)-1$ becomes the zero-map on $\mathcal{NS}$, hence on all of $\mathsf A$,
since $\mathcal{NS}$ affinely spans $\mathsf A$  \textendash\ as the local models already  do~\cite{pironio2005}. As $p\mapsto\sum_{v\in e}p(v)-1$ vanishes on $\mathcal{NS}$ \textendash\ which is by definition the solution set of $\mathrm N_{\bm x}(p)=\mathrm{NS}_i(p)=0\ \forall \bm x,i, \forall p\geq0$ \textendash\ it is a linear combination of these equations. Hence Eq.~\eqref{eq:lincomb} holds.

Thus $e$ is a valid disjunctive FR hyperedge.
\end{proof}

As we now show, Eq.~\eqref{eq:NS_normalized} is precisely the structure of pairwise exclusive completeness\footnote{ Note that Lemma~\ref{lem:valid_iff_NS_normalized} also has a consequence for the structure of the FR hierarchy. Beyond the disjunctive FR product lies the \emph{completion FR product}
(Appendix~\ref{appendix:virtual_hyperedges}, Def.~\ref{def:completion_FR_product}),
the largest FR product;  obtained by adding, if any, the \emph{virtual hyperedges}~\cite{sainz2018multipartite}. For the composition of unipartite Bell scenarios, Lemma~\ref{lem:valid_iff_NS_normalized} implies that there are none, proving a special case of Conjecture 2 in \cite{sainz2018multipartite} (cf. Appendix~\ref{appendix:virtual_hyperedges}, Corollary~\ref{cor:no_virtual_hyperedges}). }.\medskip

\begin{theorem}\label{thm:uap_pf}
  There is a one-to-one correspondence between hyperedges of the disjunctive FR product of unipartite Bell scenarios (Def.~\ref{def:disjunctive_FR_product_wo_rank})  and process functions (Def.~\ref{def:mp_pfs}).  
\end{theorem}
\begin{proof}
Throughout, $p_{\bm f}$ denotes the local deterministic model
\eqref{eq:local_det_model} associated with a tuple of response functions
$\bm f=(f_1,\dots,f_n)\in\mathcal F:=\bigtimes_{k=1}^n\mathcal A_k^{\mathcal X_k}$, where $\A_k^{\mathcal \X_k}$ denotes the set of functions $f_k:\X_k\rightarrow\A_k$, and we
abbreviate $|\mathcal A|=\prod_{k=1}^n|\mathcal A_k|$.

\medskip\noindent
($\Rightarrow$) Let $e$ be a hyperedge of
$\prescript{\mathrm{disj}}{}{\otimes}_{k=1}^n B_{1,m_k,d_k}$. By definition of the
disjunctive FR product, $e$ is a clique of the orthogonality graph, i.e.\ a set of
pairwise exclusive events (Observation~\ref{obs:exclusivitybell}). Let us show that it is also complete. 
By  Lemma~\ref{lem:valid_iff_NS_normalized} the
normalization Eq.~\eqref{eq:NS_normalized} holds for every no-signaling model, in particular
for the \emph{constant} local deterministic models $p_{\bm f^{\bm a}}$,
$\bm a\in\mathcal A$, defined by $f^{a_k}_k(x_k)=a_k$ for all $x_k\in\mathcal X_k$: each
party $k$ outputs $a_k$ irrespective of its setting. Such a model assigns probability $1$
to the events of $e$ with outcome tuple $\bm a$ and $0$ to all others, so that for all $p_{\bm f^{\bm a}}$,
   
   $1=\sum_{v:=(\bm a'|\bm x')\in e}p_{\bm f^{\bm a}}(\bm a'|\bm x')=\sum_{(\bm a'|\bm x')\in e}\delta_{\bm a',\bm a},$
   
Every outcome tuple $\bm a\in\mathcal A$ thus occurs in exactly one event of $e$. Thus
$|e|=|\mathcal A|$ and $e=\{(\bm a|\bm w(\bm a))\}_{\bm a\in\mathcal A}$ for a unique
map $\bm w:\mathcal A\to\mathcal X$. Being complete and pairwise exclusive, $e$ is
therefore the event list of a process function $\bm w$ by
Theorem~\ref{thm:characPE}.

\medskip\noindent
($\Leftarrow$) Let $\bm w$ be a process function, defining the complete pairwise exclusive
event list $\mathcal S_{\bm w}=\{(\bm a|\bm w(\bm a))\}_{\bm a\in\mathcal A}$. By
Ref.~\cite{dourdent2025paradox} (Section VII B), its mutually exclusive events obey a trivial Bell
inequality~\cite{augusiak2011bell, augusiak2012tight}, i.e.\ an inequality not
violated by any no-signaling correlation \textendash\ equivalently, a trivial local orthogonality
inequality~\cite{fritz2013local}:
\begin{equation}\label{eq:LO_trivial}
  \sum_{v\in\mathcal S_{\bm w}}p(v)\le 1
  \qquad\text{for every no-signaling model } p .
\end{equation}
It remains to show that Eq.~\eqref{eq:LO_trivial} is in fact saturated by every
no-signaling model, since Lemma~\ref{lem:valid_iff_NS_normalized} then identifies
$\mathcal S_{\bm w}$ as a hyperedge of
$\prescript{\mathrm{disj}}{}{\otimes}_{k=1}^n B_{1,m_k,d_k}$.

Averaging the local deterministic models defined in  Eq.~\eqref{eq:local_det_model} over all $\bm f\in\mathcal F$ yields the uniform
model, 
\begin{equation*}
  \frac{1}{|\mathcal F|}\sum_{\bm f\in\mathcal F}p_{\bm f}=p_\ast ,
  \qquad p_\ast(\bm a|\bm w(\a))=\frac{1}{|\mathcal A|} .
\end{equation*}
Since $\mathcal S_{\bm w}$ is complete, $|\mathcal S_{\bm w}|=|\mathcal A|$ and therefore
$\sum_{v\in\mathcal S_{\bm w}}p_\ast(v)=1$. As the sum over $\mathcal S_{\bm w}$ is linear
in the model,
\begin{equation*}
  \frac{1}{|\mathcal F|}\sum_{\bm f\in\mathcal F}
  \ \sum_{v\in\mathcal S_{\bm w}}p_{\bm f}(v)
  =\sum_{v\in\mathcal S_{\bm w}}p_\ast(v)=1 ,
\end{equation*}
while $\sum_{v\in\mathcal S_{\bm w}}p_{\bm f}(v)\le 1$ for every $\bm f$ by
Eq.~\eqref{eq:LO_trivial}. An average of numbers bounded by $1$ can equal $1$ only if each of
them does, so $\sum_{v\in\mathcal S_{\bm w}}p_{\bm f}(v)=1$ for every local deterministic
model. As these affinely span\footnote{This is not in tension with the strict inclusion
of the local polytope in the no-signaling one: the two are separated by inequalities (Bell inequalities), not by equalities, and therefore have the same affine hull~\cite{pironio2005}. Affine \emph{identities} valid on the local polytope consequently extend to the no-signaling polytope, whereas inequalities do not.} the subspace $\mathsf A$ defined by the normalization and
no-signaling conditions~\cite{pironio2005}, the affine map
$p\mapsto\sum_{v\in\mathcal S_{\bm w}}p(v)-1$ vanishes on all of $\mathsf A$, hence in
particular on every no-signaling model. Thus $\mathcal S_{\bm w}$ satisfies
Eq.~\eqref{eq:NS_normalized}, and Lemma~\ref{lem:valid_iff_NS_normalized} shows that it is a
hyperedge of $\prescript{\mathrm{disj}}{}{\otimes}_{k=1}^n B_{1,m_k,d_k}$.

\end{proof}

In the following sections, we explore several implications of Theorem~\ref{thm:uap_pf}.

\section{Applications}\label{sec:applications}

\subsection{No-signaling based characterisation of process functions}
\label{sec:ns_characterization}

Theorem~\ref{thm:uap_pf}  unveils a  connection between two seemingly separate concepts: process functions and measurement contexts.  Specifically, a deterministic classical process given by 
\begin{align}
    P(\bm{x}|\bm{a}) = 
    \begin{cases}
        1 \quad \bm{x} = \bm{w}(\bm{a}) \\
        0 \quad \text{otherwise}
    \end{cases}
\end{align}
 determines a measurement context $e$ consisting of  events $\{(\bm{a}|\bm{x})\ |\ P(\bm{x}|\bm{a})=1\}$. In particular, the properties of $e$ and the properties of process functions are equivalent:   
\begin{corollary}
    Logical consistency of process functions (Def.~\ref{def:mp_pfs}) is equivalent to the validity criterion of a  hyperedge of 
    the disjunctive FR product of unipartite Bell scenarios (Def.~\ref{def:disjunctive_FR_product_wo_rank}). \label{cor:eq_validity_criteria}
\end{corollary}

Theorem~\ref{thm:uap_pf} then implies a new necessary and sufficient criterion for the logical consistency of process functions. Whereas the characterization of Theorem~\ref{thm:characPE} (\cite{dourdent2026makescausalloopconsistent}) rests on completeness and pairwise
exclusivity, the following is phrased entirely in terms of no-signaling and normalization conditions.

\begin{corollary}   A  function $\bm{w}:\mathcal{A}\rightarrow\mathcal{X}$ is  a process function if and only if the normalization of its event set  $e:=\{(\bm{a}|\bm{w}(\bm{a}))\}$, namely $\sum_{v\in e}p(v)=1$, is a linear  combination of the no-signaling and normalization conditions  (cf. Section~\ref{sec:disj_frp}, paragraph $c$);  an  explicit matrix-based  criterion is given in Corollary~\ref{cor:ns_characterization_w_rank}.  \label{cor:ns_characterization}\end{corollary}

\subsection{Identifying Foulis-Randall products with process function types  \label{section:hierarchy_FR}}

\begin{figure*}[ht]
	\begin{center}
	\includegraphics[width=.7\textwidth]{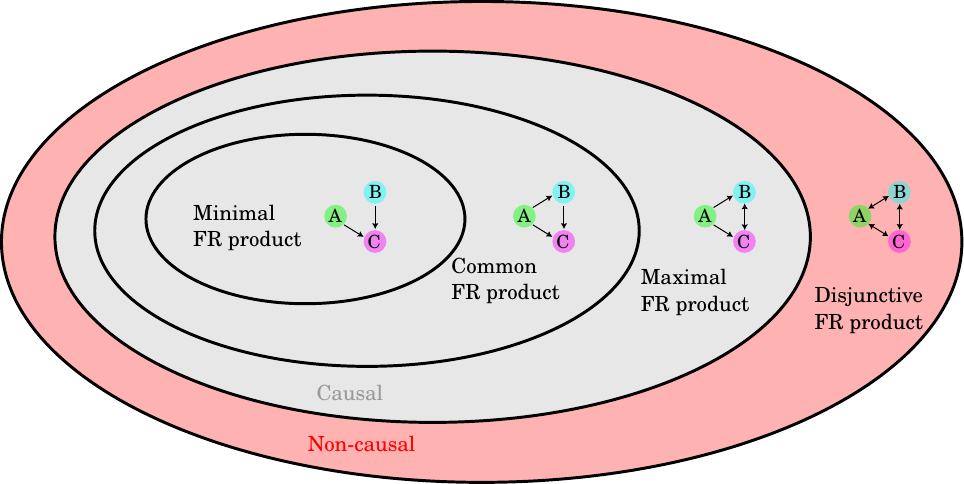}
	\end{center}
	\caption{A hierarchy of process functions from multipartite Foulis-Randall (FR) products.  
    The $n$-partite composite contextuality scenarios are constructed from the FR products of $n$ unipartite Bell scenarios.
    The process functions defined via the minimal, common, and maximal FR product are causal. 
    The disjunctive FR product also contains  hyperedges that correspond to non-causal process functions, e.g., the three-partite Lugano process (cf. Eq.~\eqref{eq:luganob}). 
    The innermost layer is defined by the minimal FR product, corresponding  to ``all-to-one" process functions, where all parties are communicating to one party.  
    The underlying causal structure is depicted by a causal graph (nodes are the parties, directed edges are the causal connections or rather, the functional dependencies between inputs and outputs of the corresponding process function). 
    The next level contains process functions defined by the common FR product. These show a type of ``causal memory" in the sense that the causal order is predefined and each party chooses a measurement based on the outcomes of the parties that precede in the measurement  protocol. 
    Next, the maximal FR product encompasses all causal process functions, while allowing the possibility of one  party controlling the order of the  other parties  based on its measurement outcome (``dynamical control" of causal order).  Finally, the disjunctive FR product includes also non-causal process functions, i.e., the possibility of two-way communication between all parties. By determining if a given process function is contained in the  disjunctive but not the maximal FR product, one can decide if a process function is causal or not. }
    \label{fig:hierarchy_FR_PF}
\end{figure*}

Recall that there is no unique way to compose more than two scenarios, leading to a multitude of FR products (as discussed at the beginning of Section~\ref{sec:equ_pf_hedges}). 
Since the largest of these products already corresponds to process functions 
(Theorem~\ref{thm:uap_pf}), the same holds for every
product in the chain. 
\begin{obs}
    For Bell scenarios $H_k=B_{1,m_k,d_k}$, the hyperedges of the minimal, common,
    and maximal FR products are process functions.
    \label{cor:hedges_min_com_max}
\end{obs}
In particular, the nesting in Eq.~\eqref{eq:fr_hierarchy} induces a hierarchy of process functions: different FR products correspond to distinct types of process functions. 
Refs.~\cite{acin2015combinatorial, sainz2018multipartite} already offered an intuition for the first three products, reading their hyperedges as parties acting within  \emph{measurement protocols} (the maximal product entailing all of them). The  disjunctive FR product, however, contains precisely the additional valid hyperedges that no such
protocol can account for. Our correspondence to process functions sharpens this picture into a genuine causal interpretation: each product in the chain of Eq.~\eqref{eq:fr_hierarchy} corresponds to a distinct class of causal structures,
ranging from process functions with fixed causal order to dynamical control of causal order, and finally
to non-causality for the disjunctive FR product. We analyze the four
levels in turn (Fig.~\ref{fig:hierarchy_FR_PF}). Moreover, we compare each level with an existing hierarchy for quantum circuits \cite{wechs2021,mothe2025}.

\subsubsection{Minimal Foulis-Randall product}
This FR product imposes an ``all-for-one" communication structure, where $(n-1)$ non-communicating parties all send a signal to the remaining party laying in their global future (central part of Fig.~\ref{fig:hierarchy_FR_PF}). 

The edge set is defined in analogy to  the bipartite FR product (cf. Eq.~\eqref{eq:hyperedges_bipartite_FR}) \cite{sainz2018multipartite}:
\begin{equation}
    E(^{\text{min}}\otimes_{k=1}^nH_k) = \bigcup_k E_k = \bigcup_k \bigcup_{\bm{v}_{\backslash k}}\{\bm{v}_{\backslash k}\}\times f(\bm{v}_{\backslash k}),
    \label{eq:fr_min_2}
\end{equation}
i.e., the set of hyperedges  is the union of sets of measurements $E_k$,
where each party $i\neq k$ conducts  a measurement $e_i$, announces the outcomes $v_i$ to a special party $k$, which then conducts the measurement as a function of the announced outcomes, i.e., $e_k = f(\bm{v}_{\backslash k})$.

Considering the case of local unipartite Bell scenarios, the process functions corresponding to the hyperedges defined in  Eq.~\eqref{eq:fr_min_2} are of the general form
\begin{equation}
    \bm{w}(\bm{a})=(c_1, c_2, ..., c_{k-1}, w_k(\bm{a}_{\backslash k}), c_{k+1}, ... ,c_n)
\end{equation}
for constants $c_i\in \X_i, i\neq k$, i.e., all components are constant except the $k$th, which can depend on all outputs except its own.
\medskip

In the quantum circuit hierarchy \cite{wechs2021}, up to a relabeling of the last party, this class of process functions can be read as the deterministic and classical version of quantum circuit with operations used in parallel (QC-PAR). We prove it explicitly in Appendix~\ref{appendix:QCQC}. 

\subsubsection{Common Foulis-Randall product}

As remarked in \cite{acin2015combinatorial}, the definition of the minimal product arbitrarily singles out one party to conduct its measurement as a function of all the others. 
One may instead consider so-called correlated measurements~\cite{sainz2018multipartite}:
 one party initiates the protocol, chooses  a measurement and  communicates its outcome to all the other parties. The remaining parties then proceed in turn, each choosing a setting that may depend on the outcomes of all the parties preceding it. One may describe this protocol as imposing a  ``causal memory'' (cf. Fig.~\ref{fig:hierarchy_FR_PF}, second innermost layer). 
 The edge set of the common FR product contains all possible correlated measurements.

Considering the case of local unipartite Bell scenarios, the process functions corresponding to the hyperedges of the common FR product take the general form  
\begin{equation}
    \bm{w}(\bm{a})=(c_1,w_2(a_1), w_3(a_1,a_2),...,w_n(\bm{a}_{\backslash n}))
\end{equation}
where $c_1\in\X_1$ is a constant, corresponding to the case where party 1 initiates the protocol.
\medskip
In the quantum circuit hierarchy \cite{wechs2021}, this class of process functions corresponds to the deterministic and classical version of quantum circuit with fixed causal order (QC-FO), also known as quantum combs \cite{chiribella2008,chiribella2009}. We prove it explicitly in Appendix~\ref{appendix:QCQC}.

\subsubsection{Maximal Foulis-Randall product}
So far, the order of the parties in the protocol is fixed. The maximal FR product extends it to any measurement protocol compatible with a well-defined causal structure. This notably includes ``dynamical'' control of causal orders \cite{mothe2025}  the causal order between parties can be controlled by the measurement outcome \textendash\ and thus, as ($f_k(x_k):=a_k$), the choice of operation  \textendash\ of another party. 
The canonical example is the classical switch, e.g. $x_1:=0,\ x_2:=a_1a_3,\ x_3:=(a_1\oplus 1)a_3$. Here, the measurement  outcome $a_1$ for setting $x_1:=0$ of party $A_1$ controls the order of the other two parties: if $a_1=0$, $A_2$ precedes $A_3$, and if $a_1=1$, $A_3$ precedes $A_2$. 

\begin{figure}[ht]
	\begin{center}
	\includegraphics[width=.45\textwidth]{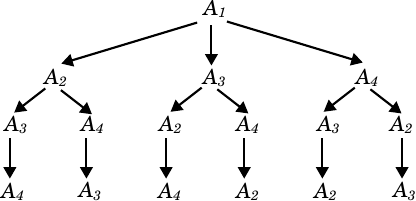}
	\end{center}
	\caption{Illustration of a measurement protocol.  Considering four parties $A_1$ to $A_4$ with $A_1$ initiating the protocol, the outcomes of the measurement chosen by $A_1$ determine sub-protocols, creating leaves and eventually, due to the recursive nature of Def.~\ref{def:measurement_protocol}, an entire directed tree. This measurement protocol is an example of dynamical control of definite causal order.
    }
    \label{fig:measurement_protocol}
\end{figure}

To define the edge set of the  maximal FR product, \cite{acin2015combinatorial} introduced a recursive definition of a  measurement protocol:
\begin{definition}[Measurement protocol~\cite{acin2015combinatorial}]
A measurement protocol $\mathcal{P}$ for a set of parties $S$ consists of the following data: 
\begin{itemize}
    \item if $S=\{\varnothing\}$, the unique protocol is $\mathcal{P} = \varnothing$
    \item otherwise, the protocol is a triple $\mathcal{P} = (k, e, f)$ for some party $k\in S$, a hyperedge $e\in E(H_k)$, and a function $f$ assigning to each $v\in e$ a measurement protocol $f(v)$ on $S\backslash \{k\}$
\end{itemize}
\label{def:measurement_protocol}
\end{definition}
\noindent As described in \cite{acin2015combinatorial, sainz2018multipartite}, this definition can be understood as 
one party $k$ initiating the protocol, choosing an initial measurement $e$, and using a function  $f$ to assign to every outcome of $e$ another measurement protocol that only includes the other parties. 
For instance, if the outcome of the initial measurement is  $v_1$, then party $l$ may be the next initiator, but if the outcome is $v_2\neq v_1$, a different party $m$ may be the next initiator \textendash\ or the same, reproducing the fixed order-case of the common FR product.  
Each of these subsequent initiators can again control which party follows.   One may think of the measurement protocol as a tree where the leaves are created by the different outcomes of the initiating parties (cf. Fig.~\ref{fig:measurement_protocol}).
Within one measurement, one can thus have various orders of parties
(cf. third layer in Fig.~\ref{fig:measurement_protocol}). Note that  every measurement constructed in this way has again a global past, the initiating party, while the global future is in general not fixed.

 More formally, the maximal FR product has a set of edges
 \begin{equation}
     E(^\text{max}\otimes_{i=1}^n H_i) = \bigcup_{\mathcal{P}} O(\mathcal{P}).
 \end{equation}
 where $O(\mathcal{P})$ is the set of outcomes of the measurement protocol $\mathcal{P}$.

Considering the case of local Bell scenarios, if party $k=1$ initiates the protocol, the corresponding process functions take the general form 
\begin{equation}
    \bm{w}(\bm{a})=(c_1, w_2(\bm{a}_{\backslash 2}), ... , w_n(\bm{a}_{\backslash n}))
    \label{eq:pf_max}
\end{equation}
for some constant measurement setting $c_1$. No-self-signaling of the other components follows from the definition of a measurement protocol: once a party has been chosen, it cannot be selected again at a later stage. In particular, all such process functions are causal: 
\begin{proposition}
    The hyperedges of the maximal FR are equivalent to all causal process functions. \label{prop:max_causal}
\end{proposition}
\begin{proof}
    Due to Corollary~\ref{cor:hedges_min_com_max}, a hyperedge of the maximal FR product corresponds to a  process function $\bm{w}$. 
    It is also causal: by definition of a measurement protocol, one party initiates and thus has a constant measurement setting, i.e., one of the process function components is constant and condition $(i)$  in Theorem~\ref{thm:characcausalpf}  is satisfied. 
    Condition $(ii)$ is also satisfied since it mimics the structure of a measurement protocol. 
   
     Conversely, a causal process function defines events $(\bm{a}|\bm{w}(\bm{a}))$ where $(i)$ implies that there is one party in the global past, which takes the role of the first party in the measurement protocol. Condition $(ii)$ again is exactly the definition of a measurement protocol.  
\end{proof}

In the quantum circuit hierarchy \cite{wechs2021}, this class of process functions corresponds to the deterministic and classical version of quantum circuit with classical control of causal order (QC-CC). We prove it in Appendix~\ref{appendix:QCQC}. Moreover, Ref.~\cite{mothe2025} introduced the notion of quantum circuits with non-influenceable classical control of causal order (QCNICCs), defined as ``the QC-CCs for which there exists an implementation such that, at any intermediate time step, the state of the control system is independent of the choice of previously applied external operations and of the potential initial state preparation in the global past.''
We also show in (Appendix~\ref{appendix:QCQC}, Corollary~\ref{cor:noQCNICC}) that in the special case of process functions, i.e. deterministic classical QC-CC, the class of QCNICC does not exist, i.e. all deterministic classical QC-CC that do not belong to QC-FO are influenceable.

\subsubsection{Disjunctive Foulis-Randall product}
Process functions that correspond to hyperedges of the disjunctive FR product have the general form
\begin{equation}
    \bm{w}(\bm{a})=(w_1(\bm{a}_{\backslash 1}), w_2(\bm{a}_{\backslash 2}), ... , w_n(\bm{a}_{\backslash n})).
    \label{eq:pf_disj}
\end{equation}

In particular, the disjunctive FR product contains hyperedges that correspond to non-causal process functions (cf. Eq.~\eqref{eq:luganob}). We can use this property to establish a  witness of classical non-causality, which is implied by  Proposition~\ref{prop:max_causal}:   
\begin{corollary}
      If a hyperedge is contained in the disjunctive but not in the maximal FR product, it corresponds to a non-causal process function. \label{cor:causal_witness}
\end{corollary}

This class of process functions has no counterpart in the quantum circuit hierarchy \cite{wechs2021}. The largest class of that hierarchy, quantum circuits with quantum control of causal order (QC-QCs), admits no deterministic classical analogue, since the control system is itself quantum. Conversely, process functions \textendash\ the deterministic classical special case of process matrices \textendash\ are not contained in the QC-QCs either: they can generate correlations violating causal inequalities, which QC-QCs cannot \cite{wechs2021}.

\subsection{Generating process functions via no-signaling conditions\label{sec:nc_from_ns_derivation}} 

In the following, we show that the normalization and no-signaling conditions characterizing process functions can yield rules that permit to substitute the events of a hyperedge. These allow to generate non-causal process functions from no-communication ones as we illustrate with the derivation of the Lugano process.

\begin{figure*}[ht]
	\begin{center}
\includegraphics[width=.9\textwidth]{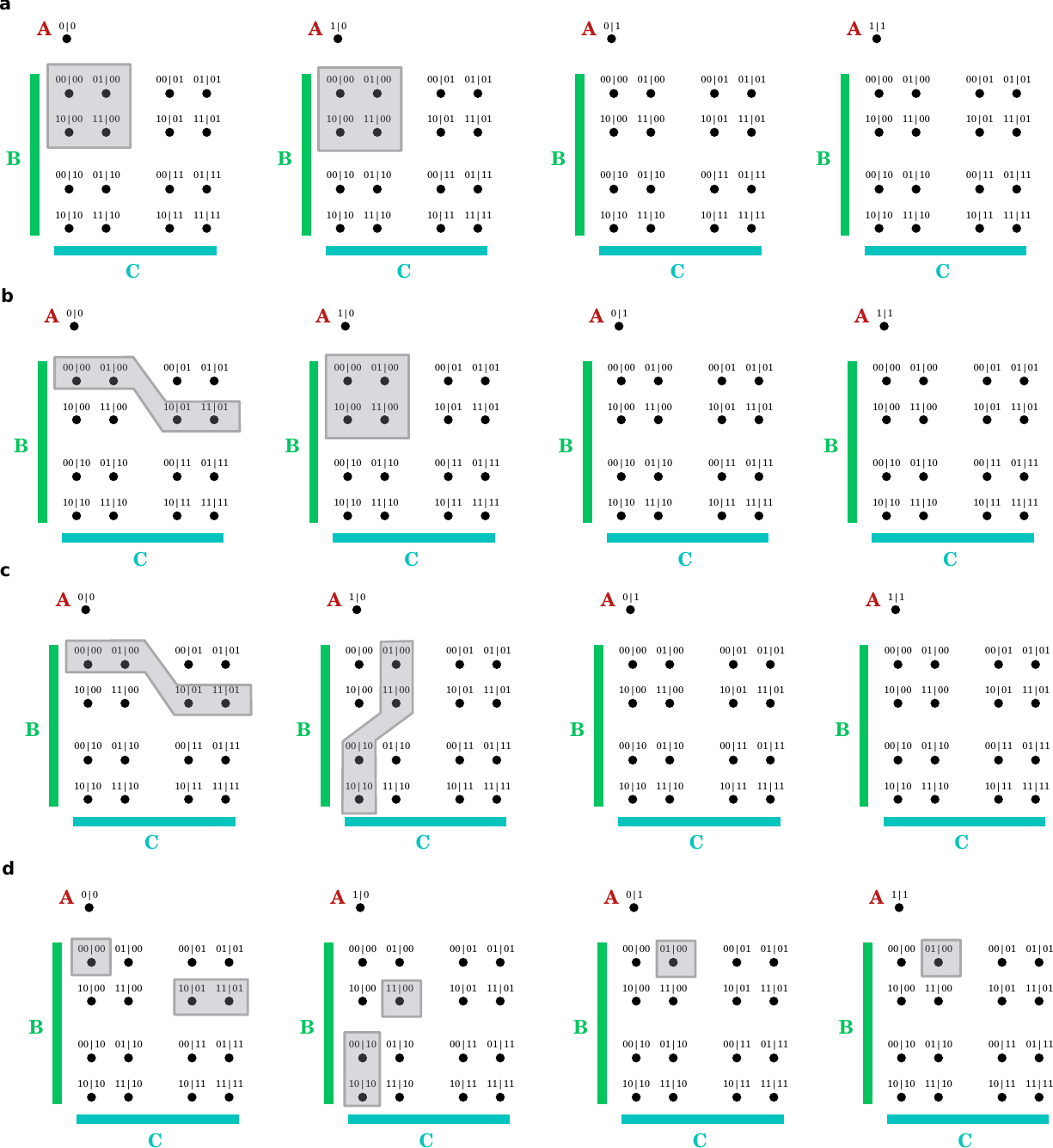}
	\end{center}
	\caption{Graphical derivation of the Lugano process. This provides a joint graphical representation of classical non-causality and non-locality without entanglement (NLWE).  
    (\textbf{a}) A three-partite Bell contextuality scenario, where for each panel, the measurement setting and outcome of $A$ is fixed. 
    The bars indicate directions in which the measurement settings and outcomes of party $B$ (green) and $C$ (cyan) vary. 
    A hyperedge with constant setting for all three parties is shown in gray.  
    (\textbf{b}) Applying a no-signaling substitution rule  (corresponding to a condition of no-signaling from $C$ to $A$ and $B$) yields a process function with communication from $A$ and $B$ to $C$. 
    (\textbf{c}) Applying another no-signaling substitution rule (corresponding to a condition of no-signaling from $B$ to $A$ and $C$) yields a process function where $A$ controls the order of $B$ and $C$, i.e., we obtain a classical switch (or rather, dynamical control of causal order).
    (\textbf{d}) Finally, applying a no-signaling condition (corresponding to a condition of no-signaling from $A$ to $B$ and $C$) yields a non-causal process: the Lugano process \cite{baumeler16}. Since this hyperedge corresponds to a SHIFT-type measurement \cite{kunjwal2023trading}, which exhibits NLWE, we obtain a joint graphical representation of NLWE and classical non-causality.
    }
\label{fig:lugano_from_afls}
\end{figure*}

\begin{figure}[ht]
	\begin{center}
\includegraphics[width=.5\textwidth]{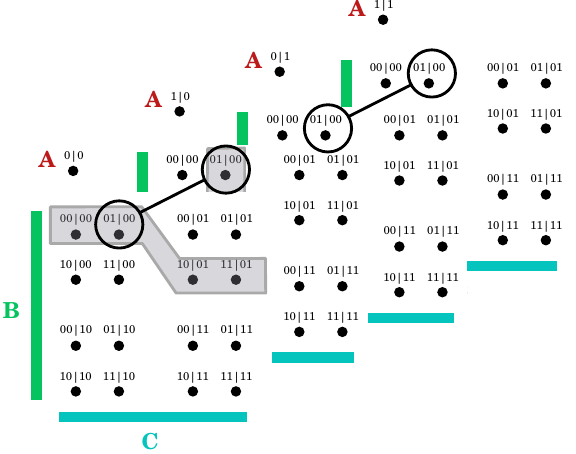}
\end{center}
\caption{Three-dimensional visualization of the no-signaling substitution rule that enables the derivation of the hyperedge corresponding to the Lugano process in Fig.~\ref{fig:lugano_from_afls}d from the hyperedge in Fig.~\ref{fig:lugano_from_afls}c.}
\label{fig:lugno_from_afls_3d}
\end{figure}
Consider the initial hyperedge that fixes every setting to $x=y=z=0$ (Fig.~\ref{fig:lugano_from_afls}a), corresponding to the
constant, no-communication, process function
\begin{equation}
    \bm{w}(\bm{a})=(w_A(b,c),\,w_B(a,c),\,w_C(a,b))=(0,0,0).
\end{equation}

The no-signaling conditions can  be  read directly off the hypergraph: in
Fig.~\ref{fig:lugano_from_afls}a, the rows encode no-signaling from $C$ to $A,B$
(the local events of $A,B$ are fixed while $C$'s setting varies), the columns encode no-signaling from $B$ to $A,C$, and events identified across panels encode no-signaling from $A$ to $B,C$ (Fig.~\ref{fig:lugno_from_afls_3d}).
More formally, the no-signaling condition preventing $C$ from signaling to $A$
and $B$ reads
\begin{equation}
    \sum_{v_C\in e_C}p(v_A, v_B, v_C) = \sum_{v_C\in e_C^\prime}p(v_A, v_B, v_C),
    \label{eq:ns_C_to_AB}
\end{equation}
for fixed $v_A\in V(H_A)$, $v_B\in V(H_B)$. Read as a substitution rule, it allows
the events on the left to be replaced by those on the right.

\medskip

\paragraph{Step 1:} 
Applying the rule for $C$ not signaling to $A,B$ (Eq.~\eqref{eq:ns_C_to_AB} with $v_A=(0|0)$, $v_B=(1|0)$
and $z=0,1$), 
\begin{equation}
    p(010|000) + p(011|000) = p(010|001) + p(011|001),
\end{equation} 
yields the hyperedge of Fig.~\ref{fig:lugano_from_afls}b—obtained by ``bending'' the
lower part of Fig.~\ref{fig:lugano_from_afls}a to the right. The process function
becomes
\begin{equation}
    \bm{w}(\bm{a})=(0,\,0,\,a\cdot b).
\end{equation}

\paragraph{Step 2:} 
Applying the rule for $B$ not signaling to $A,C$ gives
Fig.~\ref{fig:lugano_from_afls}c: $A$ now holds a fixed setting whose outcome
controls the order of $B$ and $C$ \textemdash\ a \textit{dynamical control of causal order}, 
\begin{equation}
    \bm{w}(\bm{a})=(0,\,a(c\oplus 1),\,b(a\oplus 1)).
\end{equation}

\paragraph{Step 3:} 
Finally, applying the rule for $A$ not signaling to $B,C$ (visualized
along the third dimension in Fig.~\ref{fig:lugno_from_afls_3d}) gives
Fig.~\ref{fig:lugano_from_afls}d, in which every party's setting varies. The result is the
Lugano process,
\begin{equation}
    \bm{w}(\bm{a})=(c(b\oplus 1),\,a(c\oplus 1),\,b(a\oplus 1)).
\end{equation}
Note that one could choose different no-signaling conditions (e.g., Eq.~\eqref{eq:ns_C_to_AB} for $v_A=(0|0)$, $v_B=(0|0)$ which in the first step yields the process function $\bm{w}(\bm{a})=(0,0,(a\oplus 1)\cdot(b\oplus 1))$). This, however, can  always produce non-causal process functions that are  equivalent to the Lugano process.

\begin{figure*}[ht]
	\begin{center}
\includegraphics[width=.9\textwidth]{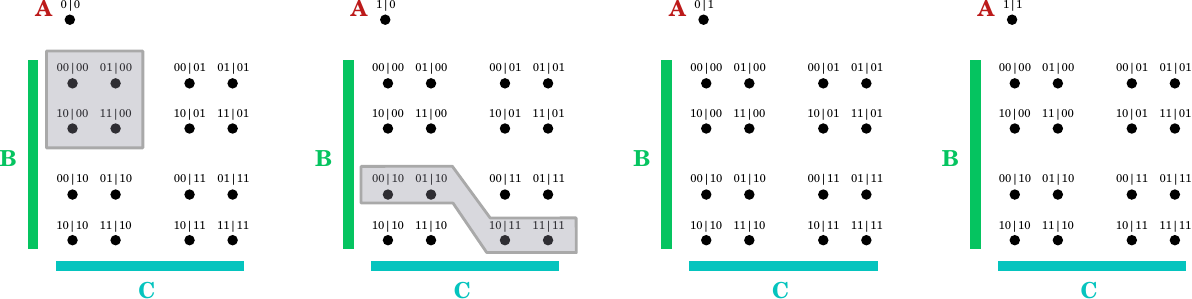
}
	\end{center}
	\caption{Example of a hyperedge that is in the common Foulis-Randall product.}
    \label{fig:common_ex}
    \end{figure*}

One can see that each step preserves logical consistency (as formalized below in Lemma~\ref{lem:pe_c_preservation}), since 
every intermediate hyperedge is itself a valid process function, tracing a path
through the causal hierarchy from a fixed order (Step~1, minimal FR product) to dynamical control
(Step~2, maximal FR product) to non-causality (Step~3, disjunctive FR product).
The only FR product not represented in this derivation is the common FR product that is in between the minimal and maximal FR product. For completeness, we provide a example for a hyperedge of the common FR  product in Fig.~\ref{fig:common_ex}. This hyperedge defines the process function
\begin{equation}
    \bm{w}(\bm{a})=(w_A(b,c),\,w_B(a,c),\,w_C(a,b))=(0,a,a\cdot b).
\end{equation}

Since all higher-dimensional non-causal process functions admit an AFLS representation, we expect that they can be derived via a similar construction.
In particular, the no-signaling conditions are defined for 
  any number of parties and any input and output
dimension. We can read these conditions as the exchange of two local settings and refer to them as \textit{no-signaling substitution rules}. 
Given the equivalence between process functions and hyperedges (Theorem~\ref{thm:uap_pf}), these rules can be read as substitutions of the hyperedge events.

Crucially, no-signaling substitution rules  preserve the defining properties of process functions and hyperedges: exclusivity and completeness. 
\begin{lemma}
    If a no-signaling substitution rule is applicable to a hyperedge of the disjunctive FR product of Bell scenarios, then it preserves exclusivity and completeness. 
    \label{lem:pe_c_preservation}
\end{lemma}
\begin{proof}
This proof is inspired by proof ideas of Conjecture 2 in \cite{sainz2018multipartite}. 
A no-signaling substitution rule exchanges $s=\{(a_k,\a_{\backslash k}|x_k, \x_{\backslash k})\}_{a_k\in\A_k}$ by $s^\prime =\{(a_k,\a_{\backslash k}|x_k^\prime, \x_{\backslash k})\}_{a_k\in\A_k}$. It can be applied to a hyperedge $e$ if  $s$ is contained in $e$. 
We show that exclusivity can never be replaced by such a no-signaling substitution rule. If it could, then exclusivity of a composite event   $\tilde{v}\in e\backslash s$ to $v\in s$ would be only due to party $k$ (if party $j\neq k$ would guarantee exclusivity between $\tilde{v}$ and $v$, it would also guarantee exclusivity for $\tilde{v}$ and $w\in s^\prime$, since $v$ and $w$ only differ for party $k$).  This would imply that $\tilde{x}_k=x_k$ and 
$\tilde{a}_k\neq a_k$.  Now, there is also an event $v^\prime \in s$  with $(\tilde{a}_k|x_k)$ since $s$ contains all $a_k\in\A_k$.
Then $\tilde{v}$ and $v^\prime$ are exclusive due to some $j\neq k$.  Since all events in $s$ share the same local events $(a_j|x_j)$ for $j\neq k$, party $j$ also makes $\tilde{v}$ exclusive with respect to $v$, contradicting the assumption that only party $k$ is responsible for this. Hence exclusivity is preserved.

Finally, completeness is preserved: a substitution rule for party $k$ exchanges $|\A_k|$  events $s$ of $e$ by  $|\A_k|$  events $s^\prime$ that are not in $e$: if  $(a_k,\a_{\backslash k}|x_k^\prime, \x_{\backslash k})\in s^\prime$ were contained in $e$, then it and $(a_k,\a_{\backslash k}|x_k, \x_{\backslash k})\in s$  would not be exclusive,  since their outputs are the same. 
\end{proof}

Since complete
pairwise-exclusive event sets are exactly process functions 
(Theorem~\ref{thm:characPE}) and hyperedges (Theorem~\ref{thm:uap_pf}), no-signaling substitution rules, when applicable, map process functions to process functions. Equivalently, they map hyperedges of the disjunctive FR product of unipartite Bell scenarios to each other.

 Note that in \cite{sainz2018multipartite} (Conjecture 2) it was mentioned that no-signaling conditions can be understood as substitution rules to generate so-called virtual hyperedges (cf. Appendix~\ref{appendix:virtual_hyperedges}), which inspired the construction in this section.

\section{Discussion\label{section:discussion}}

In this work, we provide a correspondence between two a priori unrelated objects: the hyperedges generated by Foulis–Randall (FR) products in the Acín-Fritz-Leverrier-Sainz (AFLS)  hypergraph formalism for composite Bell scenarios, and process functions — deterministic, paradox-free communication protocols that may be incompatible with any definite causal order. More precisely, the disjunctive FR product generates exactly all process functions, causal and non-causal alike, for any finite number of parties, settings, and outcomes. Logical consistency of a process function and validity of a multipartite measurement context are therefore one and the same condition.

\textit{Pairwise exclusive completeness} was recently identified as a complete characterization of process functions \cite{dourdent2026makescausalloopconsistent}. It can also be read as a structural version \cite{kunjwal2019b} of Specker's principle (``if you have several questions and you can answer any two of them, then you can also answer all of them'') \cite{cabello12s,specker} in Bell scenarios or, equivalently, of the local orthogonality principle, introduced to recover the set of quantum correlations from the no-signaling
ones~\cite{fritz2013local}. Within the AFLS framework, this condition takes a particularly
simple form: two events are exclusive exactly when they are distinct vertices of a common
hyperedge of the Bell scenario $\mathcal{B}_{n,k,m}$, so that
the complete pairwise exclusive sets of events are precisely the hyperedges in a Bell scenario. The combinatorics of Bell scenarios thus encodes every logically consistent form of deterministic classical communication, causal or non-causal alike. 

This yields a causal reading of several previously introduced notions. For instance, those measurements generated by the disjunctive FR product that so far have been regarded as non-standard measurement protocols, are exactly the non-causal process functions, while each of the remaining FR
products yields the process functions compatible with a specific class of definite causal orders.
The resulting hierarchy can in turn be compared with that of quantum circuits with quantum
control of causal order (QC-QC)~\cite{wechs2021}, the disjunctive FR product being the only introduced
product without a quantum-circuit counterpart. 
The same causal reading explains why enforcing
no-signaling for composite measurements requires communication. For instance, in the bipartite case,
measurements with one-way communication must be included~\cite{barnum2005influence}:
imposing no-signaling amounts to admitting all bipartite process functions, hence all
causal orders compatible with the no-self-signaling  and no-two-way-signaling  constraints. More generally, logical consistency of deterministic classical communication is equivalent with the normalization and no-signaling constraints that define  
the disjunctive FR product.

This yields a new characterization of process functions in terms of no-signaling and normalization conditions, complementing previously established characterizations ~\cite{baumeler2019reversible,tobar2020reversible,dourdent2025paradox,dourdent2026makescausalloopconsistent}. 
The disjunctive FR product can also  be used to witness classical non-causality: if a process function corresponds to a hyperedge of the disjunctive but not the maximal FR product (which covers all causal process functions), it is non-causal. 
Furthermore, the no-signaling conditions of the disjunctive FR product yield substitution rules that allow to generate non-causal process functions from no-communication ones.

In fact, two connections related to our main result were already hinted at in Ref.~\cite{eftaxias2023multisystem}.  First, \textit{unit effects} in boxworld can be identified with classical processes; in the deterministic case these are process functions, so that process functions correspond to deterministic unit effects.  The deterministic \textit{wiring} and \textit{non-wiring} measurements of Ref.~\cite{eftaxias2023multisystem} therefore correspond to causal and non-causal process functions,
respectively, and their algorithms generate process functions. 
Second, the authors of Ref.~\cite{eftaxias2023multisystem} also noted an equivalence between deterministic unit effects
and the hyperedges of the FR product of unipartite Bell scenarios. Lemma~\ref{lem:valid_iff_NS_normalized} proves it explicitly: deterministic unit effects determine exactly the sets of events satisfying Eq.~\eqref{eq:NS_normalized}, and these are exactly the hyperedges of the disjunctive FR product. Together, these
observations suggest the link between hyperedges and process functions that we establish
rigorously here, together with the precise relation between logical consistency and the
combinatorial structure of Bell scenarios. 
In a similar vein, Ref.~\cite{tselentis2023admissible}
proposed a graph-theoretic criterion for logical consistency, the \textit{siblings-on-cycles} (SOC) property, which constrains the graph encoding of the causal order of the parties and, through the above equivalence, also identifies the valid unit
effects. Since unit effects specify the valid hyperedges of the disjunctive FR product~\cite{eftaxias2023multisystem}, the SOC property equally constrains the admissible
hyperedges.

 Our results thus add to a growing web of connections between process functions and other
objects. The equivalence between the Lugano process, a tripartite non-causal process
function, and the \textit{SHIFT measurement}, which exhibits \textit{quantum nonlocality without
entanglement} (QNLWE)~\cite{bennett1999quantum}, was first pointed out
in Ref.~\cite{kunjwal2023nonclassicality} and later generalized to arbitrary dimensions
in Ref.~\cite{dourdent2025paradox}. This generalization rests on \textit{unambiguous product bases}, which are in correspondence with all process functions~\cite{dourdent2025paradox}, and whose events generate  \textit{trivial} Bell inequalities~\cite{augusiak2011bell,augusiak2012tight}, i.e., inequalities satisfied by all no-signaling correlations. This last property is
precisely what we exploit in the proof of Theorem~\ref{thm:uap_pf}, where we show that such inequalities are moreover saturated by every no-signaling model \textendash\ so that the events in question are exactly those generated by the disjunctive FR product of unipartite Bell scenarios. We thereby obtain the first joint hypergraph representation of non-causality and an event-based notion of NLWE \textendash\ one that the no-signaling substitution rules make straightforward to construct. Indeed, our result highlights that these phenomena are not specific to quantum theory. This resonates with previous appearances of a  SHIFT-type measurement within non-quantum theories, i.e. generalized probabilistic theories (GPTs)~\cite{hardy2001quantum, barrett2007information, plavala2023general}: it was first identified in boxworld
in Ref.~\cite{short2010strong} (proof of Theorem~3 and Fig. 2), and later rediscovered in Ref.~\cite{eftaxias2023multisystem} (Section VII B) where it was explicitly presented as a deterministic non-wiring, which we identify as a Lugano-like process function. It was also shown by
Pusey~\cite{pusey2012stabilizer} to arise in Spekkens' toy theory~\cite{spekkens2007evidence}, which reproduces many quantum phenomena except  nonlocality and contextuality.  Our work thus contributes to a recent line of research on indefinite causal order beyond quantum theory~\cite{chen2024causality, sengupta2024achieving,
bavaresco2024indefinite,liu2025,liu2026}.

In a different direction, although process functions are usually presented as the classical version of quantum indefinite causal order~\cite{oreshkov2012quantum,costa2026indefinitequantumcausality}, they have been given some alternative readings. They have been related to agent-environment interactions in artificial
intelligence~\cite{wilson2025agent}, and to a recent interpretation of indefinite causal
order in which agents are groupings of input and output labels (``agents from
bits''~\cite{grinbaum2026}). In line with the latter, we propose the notions of \textit{agent} and \textit{process events}, which encode
two complementary perspectives: the one of the local agents performing operations and the one of the global
superobserver represented by the process itself (see Appendix~\ref{sec:agency}). In the same spirit, and following the relation established with Ref.~\cite{eftaxias2023multisystem}, our results recast process functions \textendash\ non-causal ones included \textendash\ as the logically consistent deterministic compositions of black boxes.
\medskip

Finally, a natural question for future work is whether the causal hierarchy of FR products has observational consequences. For classical, quantum, and boxworld models it does not: the correlations are insensitive to which FR product is used \cite{acin2015combinatorial,sainz2018multipartite}, so that the richer communication structures \textendash\ non-causal process functions in particular \textendash\ leave no observational trace in this context. Other theories are more discriminating: $Q_1$ models (\cite{acin2015combinatorial}, Prop. 6.3.1) distinguish the minimal product from the rest, yielding \textit{almost quantum correlations} \cite{navascues2015almost} beyond it. This raises the question whether some alternative theory is \emph{not} observationally equivalent for the
maximal and the disjunctive FR product. In such a theory, including the non-causal measurements in a Bell scenario would change its no-signaling correlations, so that classical non-causality would acquire a new kind of signature which may offer a new handle on it, or an argument
for ruling it out.

\medskip

\begin{acknowledgments}
We thank  Mirjam Weilenmann for enlightening discussions about the connections between unit effects, classical processes, and hyperedges via e-mail  and at Causalworlds 2026. 
Moreover, we thank Ana Belén Sainz and Elie Wolfe for helpful discussions. We acknowledge the use of the 
AI assistant Claude Opus 5.5 (Anthropic). Interaction with it helped complete the proof of Theorem~\ref{thm:uap_pf}, in particular through Lemma~\ref{lem:valid_iff_NS_normalized}. It was further used to check the proofs of  Lemma~\ref{lem:pe_c_preservation} and Corollary~\ref{cor:no_virtual_hyperedges}.  The assistant was finally used to
improve the English of the manuscript. All mathematical statements and proofs were independently derived, checked, and are the sole responsibility of the authors.
H.D., A.L., and A.A. acknowledge financial support from the Government of Spain (Severo Ochoa CEX2019-000910-S, NextGenerationEU PRTR-C17.I1 and FUNQIP), Fundació Cellex, Fundació Mir-Puig, Generalitat de Catalunya (CERCA program). A.A was also supported by the ERC AdG CERQUTE, the AXA Chair in Quantum Information Science. 
K.S. and A.L. acknowledge: This research was funded in whole or in part by the Austrian Science Fund (FWF) 10.55776/PAT4559623. For open access purposes, the author has applied a CC BY public copyright license to any author-accepted manuscript version arising from this submission. 
A.L.\ acknowledges the European Union (PASQuanS2.1, 101113690). 
A.L.\ benefited from the activities of COST Action CA23115: Relativistic Quantum Information, funded  by COST (European Cooperation in Science and Technology), specifically discussions during the conference \emph{Observers and Causality in Quantum Gravity} in Bratislava, Slovakia (2026).
R.K. acknowledges: This work received support from the French government under the France 2030 investment plan, as part of the Initiative d’Excellence d’Aix-Marseille Université-A*MIDEX, AMX-22-CEI-01.
Views and opinions expressed are however those of the author(s) only and do not necessarily reflect those of the European Union or the European Commission. Neither the European Union nor the granting authority can be held responsible for them. 
\end{acknowledgments}

\bibliography{references}

\newpage

\appendix

\section{Virtual hyperedges\label{appendix:virtual_hyperedges}}

A contextuality scenario $H$ may allow for additional hyperedges: 
\begin{definition}[Virtual edges]
    A probabilistic model on a contextuality scenario $H$ can allow for  virtual hyperedges that are not contained in $H$ but satisfy the normalization condition (Eq.~\eqref{eq:normalization}).
    \label{def:virtual}
\end{definition}
\noindent Intuitively, one may consider virtual hyperedges as those sets of vertices that ``want to be" a measurement~\cite{acin2015combinatorial}. 

To illustrate this definition, consider the pentagon scenario shown in Fig.~\ref{fig:AFLS_introduction}b whose normalization conditions are  
\begin{align*}
    p(v_1)+p(v_2)&=1 \quad(\text{I})\\
    p(v_2)+p(v_3)&=1\quad(\text{II})\\
    p(v_3)+p(v_4)&=1\quad(\text{III})\\
    p(v_4)+p(v_5)&=1\quad(\text{IV})\\
    p(v_5)+p(v_1)&=1\quad(\text{V}).
\end{align*}
Subtracting (IV) from (III) 
yields
\begin{align*}
    p(v_3)-p(v_5)=0.
\end{align*}
 Adding (V) to this equation gives
\begin{equation*}
    p(v_3)+p(v_1)=1,
\end{equation*}
which shows that the dashed hyperedge in Fig.~\ref{fig:AFLS_introduction}b (bottom) is indeed normalized and thus constitutes a virtual hyperedge.

Contextuality scenarios such as the above example are called \emph{incomplete}, while one may construct a completion of it:
\begin{definition}[Completion]
    The completion $\overline{H}$ of a contextuality scenario $H$ is obtained by adding all the virtual edges to $H$.
    \label{def:completion}
\end{definition}
\noindent This allows to define a multipartite FR product that contains all other ones: 
\begin{definition}[Def. 10 in \cite{sainz2018multipartite}]
    The completion product is defined as the completion of any other multipartite FR product:
    \begin{align*}
        \bar{\otimes}_{i=1}^n H_i &= \overline{\prescript{\text{}}{}{\otimes_{i=1}H_i}}. 
    \end{align*} 
    \label{def:completion_FR_product}
\end{definition}

\noindent This product is in general larger than the disjunctive FR product. In this work, we consider unipartite Bell scenarios, for which the equivalence between deterministic unit effects and hyperedges of the disjunctive FR product of Bell scenarios forbids the existence of virtual hyperedges (addressing Conjecture~2 of \cite{sainz2018multipartite} for the case of unipartite Bell scenarios).  In fact, this follows immediately from Lemma~\ref{lem:valid_iff_NS_normalized}:
\begin{corollary}
    The disjunctive FR product of Bell scenarios has no virtual hyperedges; that is,
    it coincides with the completion FR product.
    \label{cor:no_virtual_hyperedges}
\end{corollary}
\begin{proof}
    Assume a virtual hyperedge $e_V$ exists. In general, its vertices may not satisfy pairwise exclusivity. Its normalization condition 
    is implied by the normalization conditions of the hyperedges of $\prescript{\mathrm{disj}}{}{\otimes}_{k=1}^n B_{1,m_k,d_k}$ and thus $\sum_{v\in e_V} p(v)=1$ holds for any $p$ defined over $\prescript{\mathrm{disj}}{}{\otimes}_{k=1}^n B_{1,m_k,d_k}$. By the first direction of Lemma~\ref{lem:valid_iff_NS_normalized} this identity holds for any no-signaling model, i.e., Eq.~\eqref{eq:NS_normalized} is satisfied by $e_V$. By the reverse direction of Lemma~\ref{lem:valid_iff_NS_normalized}, any set of vertices satisfying Eq.~\eqref{eq:NS_normalized}  is already a hyperedge of $\prescript{\mathrm{disj}}{}{\otimes}_{k=1}^n B_{1,m_k,d_k}$. Thus $e_V$ is not a virtual hyperedge. 
\end{proof}

\section{Matrix representation of the disjunctive Foulis-Randall product\label{appendix:matrix_representation}}

Following \cite{sainz2018multipartite}, one can formalize  the requirements of  normalization and no-signaling using matrix representations of hypergraphs.  
\begin{definition}[Matrix representation of a hypergraph]
    The matrix  $M$ of a hypergraph $H$ has rows that are labeled by the hyperedges $e\in E(H)$ and columns that are labeled by the vertices $v\in V(H)$, i.e., $M_{ev}$ corresponds to a potential outcome $v$ of measurement $e$. If outcome $v$  appears in $e$, $M_{ev}=1$, and if not, $M_{ev}=0$.
\end{definition} 
\noindent Normalization of probabilistic models can be expressed the following way:
\begin{definition}[Matrix representation of a  normalization condition]
    The normalization condition for a probabilistic model $p$ defined over a hypergraph $H$  can be written as $M\cdot\bm{p} = \bm{1}$ where the components of $\bm{p}$ are the probabilities $p(v),v\in V(H)$ and $[\bm{1}]_i=1\ \forall i=1,...,|E(H)|$. 
\end{definition}
\noindent To enforce that the hyperedges defining $M$ satisfy all no-signaling and  normalization conditions, the rank of the hypergraph matrix $M$ has to be equal to the rank of another matrix whose rows are defined by no-signaling and normalization    conditions. The latter two are  defined in the following.
\begin{definition}[Composite normalization matrix]
Given a composite contextuality scenario $H=\otimes_{i=1}^n H_i$, the composite normalization matrix $M_{\textit{Norm}}$ has $\{ 0,1\}$-valued entries, its columns are labeled by the vertices and its rows correspond to no-communication hyperedges: these  correspond to each scenario $k$ choosing a local measurement $e_k\in H_k$  independently of the other scenarios. $M_{\textit{Norm}}$ represents all possible combinations of such local choices and satisfies
\begin{equation}
    M_{\textit{Norm}}\cdot\bm{p}=\bm{1}.
\end{equation} 
\label{def:composite_normalization_matrix}
\end{definition}

One may rewrite the no-signaling conditions as 
  \begin{multline}
        \sum_{v_k\in e_k}p(v_1, ..., v_k, ..., v_n) - \sum_{v_k\in e_k^\prime} p(v_1,...,v_k,...,v_n) = 0 \\ \forall e_k,e_k^\prime \in H_k, \forall v_1,...v_{k-1},v_{k+1},...v_n,
        \label{eq:ns_equalities_reformulated}
    \end{multline}
which allows to define the following matrix:
\begin{definition}[Composite no-signaling matrix]
 We denote     $M_{\mathrm{NS}}$ as the matrix which represents all possible  no-signaling equalities as 
 defined in Eq.~\eqref{eq:ns_equalities_reformulated}. The rows of  $M_{\mathrm{NS}}$ correspond to a specific no-signaling condition, while the columns are labelled by composite events $v:=(v_1,...,v_n)$. A value of $+1$ is assigned to a matrix element if the column label contains $v_k\in e_k\backslash e_k^\prime$, $-1$ if $v_k\in e_k^\prime\backslash e_k$ and $0$ for all other cases.
 Thus, Eq.~\ref{eq:ns_equalities_reformulated} can be written as 
 \begin{equation}
     M_{\mathrm{NS}}\cdot\bm{p}=\bm{0}
 \end{equation} where $[\bm{0}]_i=0\ \forall i=1,...,|E(H)|$. 
 \label{def:composite_ns_matrix}
\end{definition}
\noindent Given these definitions, $H$ is said to capture no-signaling and normalization if and only if the \textit{reduced row echelon} form (rref) of the  hypergraph matrix $M^{\bm{1}}$ (padded with a column of all-ones) equals the reduced row echelon form of the combined rows of normalization and no-signaling hyperedges, $M^{\bm{1}}_{\mathrm{Norm}}\cup M^{\bm{0}}_{\mathrm{NS}}$ (padded with a column of all-ones and all-zeros, respectively):
\begin{equation}
\text{rref}\left(M^{\bm{1}}\right) = \text{rref}\left(M^{\bm{1}}_{\mathrm{Norm}}\cup M^{\bm{0}}_{\mathrm{NS}}\right)
\end{equation}

We can use these definitions to provide a matrix-based definition of the  disjunctive FR product $\prescript{\mathrm{disj}}{}{\otimes}_{k=1}^nH_k$ for $n$ local scenarios $H_k$. As discussed in the main text,  one constructs a composite, faithful orthogonality graph from the  orthogonality graphs $\text{Ortho}(H_k)$ (cf. Def.~\ref{def:faitfhul_ortho_graph}).
Then one inspects all of its cliques, i.e., sets of events that are mutually connected. These events satisfy pairwise exclusivity. 
Considering these as candidate hyperedges, the sum of probabilities over their respective events must be one. 
If these conditions do not introduce anything beyond the no-signaling and normalization constraints, then the candidate hyperedge is included in the edge set $E(\prescript{\mathrm{disj}}{}{\otimes}_{k=1}^nH_k)$. Mathematically, this can be expressed by comparing the ranks of two suitably defined matrices. 
\begin{definition}[Disjunctive FR product]
The disjunctive FR product $\prescript{\mathrm{disj}}{}{\otimes}_{k=1}^nH_k$ is defined by vertices $V(\prescript{\mathrm{disj}}{}{\otimes}_{k=1}^nH_k):=\bigtimes_{k=1}^n V(H_k)$ and the edge set
\begin{equation}
    \begin{aligned}&E(\prescript{\mathrm{disj}}{}{\otimes}_{i=1}^nH_i) := \\ &\Bigl\{ e\ \Bigl|\ 
    e\in\emph{Cliques}(\emph{FaithfulOrthoGraph}(H_1,...,H_n)) \\  
    & \land\ \mathrm{rank}\left(M_{\mathrm{Norm}}^{\bm{1}} \cup M_{\mathrm{NS}}^{\bm{0}}\right)
    = \emph{rank}\left(e^{\bm{1}}\cup M_{\mathrm{Norm}}^{\bm{1}}\cup M_{\mathrm{NS}}^{\bm{0}}\right)\Bigr\},
\end{aligned}
\label{eq:disjunctive_FR_product}
\end{equation}
where \emph{Cliques} denotes subsets of vertices that are adjacent according to the faithful orthogonality graph \emph{FaithfulOrthoGraph}, while  $M_\mathrm{Norm}^1$ and $M_\mathrm{NS}^{\bm{0}}$ denote the matrix representation of normalization and no-signaling conditions (Def.~\ref{def:composite_normalization_matrix} and Def.~\ref{def:composite_ns_matrix})  padded with a column of ones or rather zeroes.    $e^{\bm{1}}$ is a short notation for the matrix representation of $\sum_{v\in e}p(v)=1$.
\label{def:disjunctive_FR_product}
\end{definition}

One can also reformulate the characterization of Corollary~\ref{cor:ns_characterization}:
\begin{corollary}
    A no-self-signaling function $\bm{w}:\mathcal{A}\rightarrow\mathcal{X}$ is  a process function if and only if  
    it gives rise to events $e:=\{(\bm{a}|\bm{w}(\bm{a}))\}$ that satisfy 
\begin{equation}
    \mathrm{rank}\left(M_{\mathrm{Norm}}^{\bm{1}} \cup M_{\mathrm{NS}}^{\bm{0}}\right)
    = \mathrm{rank}\left(e^{\bm{1}}\cup M_{\mathrm{Norm}}^{\bm{1}}\cup M_{\mathrm{NS}}^{\bm{0}}\right)
\end{equation}
where $M_{\mathrm{Norm}}^{\bm{1}}$ and $M_{\mathrm{NS}}^{\bm{0}}$ are the matrix form of  the no-signaling and normalization conditions specified in Def.~\ref{def:composite_normalization_matrix} and \ref{def:composite_ns_matrix}
    \label{corr:characterization}. $e^{\bm{1}}$ denotes the matrix form of the normalization condition $\sum_{v\in e} p(v)$ for a probabilistic model $p$. 
    \label{cor:ns_characterization_w_rank}
\end{corollary}

Finally, one may also state the definition of virtual hyperedges (cf. Appendix~\ref{appendix:virtual_hyperedges}) using matrices:
\begin{definition}
    A hyperedge $e$ is a virtual hyperedge of $H$ if and only if
    \begin{equation}
        e\notin E(H) \ \land \ \emph{rank}\left(M^{\bm{1}}(H)\cup e^{\bm{1}}\right) = \emph{rank}\left(M^{\bm{1}}(H)\right) \label{eq:vedge_hypmatrix}
    \end{equation}
    \label{def:vedge_hypmatrix}
\end{definition}

\section{Comparison with the QC-QC hierarchy}\label{appendix:QCQC}

By Theorem~4 of Ref.~\cite{baumeler22}, any process function $\w: \mathcal{A} \to \mathcal{X}$ admits a unique encoding as  a process matrix:
\begin{align}\label{eq:classicalPM}
    W^{\mathbf{A}}_{\w}=\sum_{\a} \bigotimes_k \ketbra{a_k}{a_k}^{A_{k}^{O}}\otimes\ketbra{w_k(\a_{\backslash k})}{w_k(\a_{\backslash k})}^{A_{k}^{I}}.
\end{align}
with parties input and output spaces of linear operators $\mathcal{L}(\HS^{A^{I}_k})$ and $\mathcal{L}(\HS^{A^{O}_k})$.

\subsection{Quantum circuits with operations used in parallel (QC-PAR)}

A QC-PAR \cite{wechs2021} is a circuit composed of two internal Completely-Positive-Trace-Preserving (CPTP) operations: $\M_P:\L(\HS^P)\rightarrow\L(\HS^{A^I_{\N}\alpha})$, $\HS^{A^I_{\N}\alpha}:=\HS^{A^I_{\N}}\otimes\HS^{\alpha}$, $\HS^{A^I_{\N}}:=\bigotimes_{k\in\N}\HS^{A^I_k}$ (with Choi matrix $M_P\in\L(\HS^{PA^I_{\N}\alpha})$) which takes the circuit's inputs in the global past $\HS^P$ and  outputs a state in the input Hilbert space $\HS^{A^I_{\N}}$, which in general may be correlated with an auxiliary system in some Hilbert space $\HS^{\alpha}$ ; and $\M_F:\L(\HS^{A^O_{\N}\alpha})\rightarrow\L(\HS^{F})$ (with Choi matrix $M_F\in\L(\HS^{A^O_{\N}\alpha F})$) which takes the output state of $\HS^{A^O_{\N}}$ together with the auxiliary state to the  circuit's outputs in the global future $\HS^F$.
The corresponding process matrix is 
\begin{align}
    W=M_P * M_F \in \L(\HS^{PA^{IO}_{\N}\alpha F}),
\end{align}
where $*$ is the \textit{link product}~\cite{chiribella2008,chiribella09}. For two operators $M\in\L(\mathcal{H}^{XY})$ and $N\in\L(\mathcal{H}^{YZ})$, the latter is defined as $M*N=\Tr_{Y} [(M^{T_Y}\otimes\id^{Z})(\id^{X}\otimes N)] $ where $T_Y$ denotes  partial transposition on  $Y$,
with  the operations satisfying the constraints
\begin{align}
    \Tr_{A^I_{\N}\alpha}M_P=\id^P, \\
    \hbox{and } \Tr_F(M_P * M_F)=\Tr_\alpha M_P\otimes\id^{A^O_{\N}},
    \label{eq:QCPARcharac}
\end{align}

Considering  trivial global input and output spaces \textendash\ alternatively, one could also purify the process matrix Eq.~\eqref{eq:classicalPM} following Ref.~\cite{araujo4} \textendash\ and a trivial auxiliary state, $\dim(\HS^{P})=\dim(\HS^{F})=\dim(\HS^\alpha)=1$, the constraints Eq.~\eqref{eq:QCPARcharac} becomes
\begin{align}
    \Tr_{A^I_{\N}}M_P=1, \label{eq:QCPARcharac1} \\
    \hbox{and } M_P * M_F= M_P\otimes\id^{A^O_{\N}}. \label{eq:QCPARcharac2}
\end{align}

\begin{theorem}\label{th:QPAR}
    Process matrices of ``all-for-one'' process functions generated by the minimal FR product are QC-PARs.
    \begin{proof}
       Relabel the global future party $k$ of the ``all-for-one'' process functions  as party ``$F$'': for all $k\in\N{\setminus F}$, the input state corresponds to constant functions. Tracing out the future system thus leads to a $N-1$ process matrix of the form  Eq.~\eqref{eq:classicalPM}. Because the parties' outputs are linearly independent, averaging over all of them leads to:
       \begin{align}
           \Tr_F  W^{\mathbf{A}}_{\w}= \bigotimes_k \id^{A_{k}^{O}}\otimes\ketbra{x_k}{x_k}^{A_{k}^{I}},
       \end{align}
       satisfying Eq.~\eqref{eq:QCPARcharac2} with $M_P=\bigotimes_k \ketbra{x_k}{x_k}^{A_{k}^{I}}$. The latter straightforwardly satisfies Eq.~\eqref{eq:QCPARcharac1}.
    \end{proof}   
\end{theorem}

The process
\begin{align}
    x_1:=0,\quad x_2:=0,\quad x_3:=a_1a_2,
\end{align}
is an example of a process function generate by the minimal FR product, thus implementable in a QC-PAR.

Note that this is only true because of the relabeling. If the last party is not relabeled as $F$, then ``all-for-one'' process functions are a special case of processes with fixed causal order in which all but one party are in parallel. Consequently, the minimal FR
product does not correspond exactly to QC-PAR, but is slightly more general: the product
associated with QC-PAR in the absence of relabelling is the Cartesian product, whose
hyperedges involve no communication at all. Since no analogous relabelling is used for the
other classes, this should be kept in mind when comparing the two hierarchies.

\subsection{Quantum circuits with fixed causal order (QC-FO)}

A QC-FO \cite{wechs2021}, also known as quantum comb \cite{chiribella2008,chiribella2009}, is a circuit composed of $N+1$ internal CPTP operations placed in a fixed order $(1,...,N)$. Considering a trivial global input space, and denoting the reduced matrices
\begin{align}
    W_{(n)}:=\frac{1}{\prod_{k=n}^{N}d_k^O}\Tr_{\prod_{k={n+1}}^{N}A^{IO}_kA^O_nF}W,
\end{align}
their process matrices are characterized by the following constraints:
\begin{align}
    \Tr_F W = W_{(N)} \otimes \id^{A^O_{N}},\label{eq:QCFOcharac1} \\
    \forall n\in\N\setminus N, \hspace{2mm} \Tr_{A^I_{n+1}}W_{(n+1)}=W_{(n)}\otimes \id^{A^O_{n}},\label{eq:QCFOcharac2}, \\
    \Tr_{A^I_{1}}W_{(1)}=1.\label{eq:QCFOcharac3}
\end{align}

\begin{theorem}\label{th:QCFO}
    Process matrices of process functions generated by the common FR product are QC-FOs.
    
    \begin{proof}

    Label the parties according to the fixed order $A_1\prec\cdots\prec A_N$. Then $x_1=c_1$, $x_n=w_n(a_1,\ldots,a_{n-1})$
     for $(n\geq2)$.
    Define
    \begin{align}
        W_{(n)}
        :=
        \sum_{a_1,\ldots,a_{n-1}}
        &\bigotimes_{j=1}^{n-1}
        \left(
            \ketbra{a_j}{a_j}^{A_j^O}
            \otimes
            \ketbra{x_j}{x_j}^{A_j^I}
        \right)\notag\\
        &\otimes
        \ketbra{w_n}{w_n}^{A_n^I},
    \end{align}
    where $x_j=w_j(a_1,\ldots,a_{j-1})$. Since
    $W_{(1)}=\ketbra{c_1}{c_1}^{A_1^I}$,
    we have $\Tr_{A_1^I}W_{(1)}=1$ (Eq.~\eqref{eq:QCFOcharac3}).
    Moreover, for every $n=1,\ldots,N-1$,
    \begin{align}
        \Tr_{A_{n+1}^I}W_{(n+1)}
        &=
        W_{(n)}
        \otimes
        \sum_{a_n}\ketbra{a_n}{a_n}^{A_n^O}
        =
        W_{(n)}\otimes\id^{A_n^O},
    \end{align}
    thus Eq.~\eqref{eq:QCFOcharac2} is satisfied.
    Finally, the last output cannot influence any input, and hence
    \[
        W_{\bm w}^{\bm A}
        =
        W_{(N)}
        \otimes
        \sum_{a_N}\ketbra{a_N}{a_N}^{A_N^O}
        =
        W_{(N)}\otimes\id^{A_N^O}.
    \]
    Taking $F$ to be trivial gives
    $\Tr_F W_{\bm w}^{\bm A}=W_{(N)}\otimes\id^{A_N^O}$ (Eq.~\eqref{eq:QCFOcharac1}).
    Thus all the QC-FO constraints are satisfied.
    \end{proof}   
\end{theorem}

The process
\begin{align}
    x_1:=0,\quad x_2:=a_1,\quad x_3:=a_1a_2,
\end{align}
is an example of a process function generate by the common FR product, thus implementable in a QC-FO. It is straightforward to see that its process matrix does not satisfy Eq.~\eqref{eq:QCPARcharac1}, and thus it cannot be implemented in a QC-PAR.

\subsection{Quantum circuits with classical control of causal order (QC-CC)}

A QC-CC \cite{wechs2021}, is a circuit composed of $N+1$ internal quantum instruments (collections of CP maps which sum up to a CPTP map) in which the causal order of external operations can be established dynamically, with the operations in the past determining the causal order of the operations in the future. Their process matrices are characterized by the following constraints, for all nonempty ordered subsets $(k_1,...,k_n)$ of $\N$:
\begin{align}
    &W=\sum_{(k_1,...,k_N)}W_{(k_1,...,k_N,F)},\label{eq:QCCCchara0}\\
   &\forall (k_1,...,k_N),\ \Tr_F W_{(k_1,...,k_N,F)} =W_{(k_1,...,k_N)} \otimes \id^{A^O_{N}},\label{eq:QCCCcharac1}  \\
   & \forall\ n=1,...,N-1,\ \forall (k_1,...,k_n),\notag\\ &\sum_{k_{n+1}}\Tr_{A^I_{k_{n+1}}} W_{(k_1,...,k_{n+1})} =W_{(k_1,...,k_n)} \otimes \id^{A^O_{n}},\label{eq:QCCCchara2} \\
  &  \sum_{k_1}\Tr_{A^I_{k_1}}W_{(k_1)}=1.\label{eq:QCCCchara3}
\end{align}

\begin{theorem}\label{th:QCCC}
    Process matrices of process functions generated by the maximal FR product are QC-CCs.
    \begin{proof}

A measurement protocol of the maximal FR product is a rooted decision
    tree. At every node, the protocol selects a party that has not acted
    before, and the outcome of that party determines the next node. Although this representation selects one party at each step, several parties may be causally parallel. In particular, several parties may receive constant inputs and therefore
    be causally initial. Such parallel parties can be placed in any order without introducing a causal dependence between them.

    We fix a sequentialization of the protocol. Thus, every output string $\bm a$ determines an ordered
    branch
    \begin{align}
       \pi(\bm a)=(k_1,\ldots,k_N) 
    \end{align}
    of the parties.

   Here, $k_1$ is only the first party in the chosen sequentialization; it
    need not be the unique causally initial party. If $k_j$ is parallel to
    the parties preceding it, then its input is independent of their
    outcomes. Thus, along a branch $(k_1,\ldots,k_N)$, we may
    write
    \begin{align}
        w_{k_n}(\bm a_{\backslash k_n})
        =
        w_{k_n}(a_{k_1},\ldots,a_{k_{n-1}}).
    \end{align}
    This follows directly from the recursive definition of a measurement
    protocol: once a party has been selected, it cannot be selected again.

    For an ordered subset $(k_1,\ldots,k_n)$,  
    $(a_{k_1},\ldots,a_{k_{n-1}})$ is the set of outcome histories for which the chosen sequentialization
    selects the parties $k_1,\ldots,k_n$ in this order. Define
    \begin{align}
       & W_{(k_1,\ldots,k_n)}
        :=\notag\\
       & \sum_{\mathcal (a_{k_1},\ldots,a_{k_{n-1}})}
        \bigotimes_{j=1}^{n-1}
        (
            \ketbra{a_{k_j}}{a_{k_j}}^{A_{k_j}^O}\notag\\
         &\otimes
            \ketbra{
                w_{k_j}(a_{k_1},\ldots,a_{k_{j-1}})
            }{
                w_{k_j}(a_{k_1},\ldots,a_{k_{j-1}})
            }^{A_{k_j}^I}
        )
        \notag\\
        &\otimes
        \ketbra{
            w_{k_n}(a_{k_1},\ldots,a_{k_{n-1}})
        }{
            w_{k_n}(a_{k_1},\ldots,a_{k_{n-1}})
        }^{A_{k_n}^I}.
        \label{eq:qccc-prefix}
    \end{align}
    Here and below, an operator is defined to be zero if the corresponding
    ordered subset does not occur in the chosen sequentialization.

    Let $k_1$ be the formal first party in this sequentialization. Since
    $k_1$ is causally initial, its input is a constant $c_{k_1}$, and hence $W_{(k_1)}
        =
        \ketbra{c_{k_1}}{c_{k_1}}^{A_{k_1}^I}
        $. 
 
    Therefore,
    \begin{align}
                \sum_{k_1} \Tr_{A_{k_1}^I}W_{(k)}
        =
        \Tr\!\left(
            \ketbra{c_{k_1}}{c_{k_1}}^{A_{k_1}^I}
        \right)
        =1,
    \end{align}
    which proves Eq.~\eqref{eq:QCCCchara3}.

    Consider now a prefix $(k_1,\ldots,k_n)$ with $n<N$. For each compatible
    history $(a_{k_1},\ldots,a_{k_{n-1}})$, the outcome $a_{k_n}$ determines exactly one next party
    $k_{n+1}$. Thus, as $k_{n+1}$ ranges over the parties not yet selected, the
    operators
    $W_{(k_1,\ldots,k_n,k_{n+1})}$ partition all possible extensions of the
    prefix by the outcome $a_{k_n}$. Moreover, tracing out the input of the
    next party removes its normalized input state, since
    \begin{align}
               \Tr\!\left(
            \ketbra{
                w_{k_{n+1}}(a_{k_1},\ldots,a_{k_n})
            }{
                w_{k_{n+1}}(a_{k_1},\ldots,a_{k_n})
            }^{A_{k_{n+1}}^I}
        \right)=1.
    \end{align}
     It follows that
   \begin{align}
        &\sum_{k_{n+1}}
        \Tr_{A_{k_{n+1}}^I}
        W_{(k_1,\ldots,k_n,k_{n+1})}
        \notag\\
        &\qquad =
        W_{(k_1,\ldots,k_n)}
        \otimes
        \sum_{a_{k_n}}
        \ketbra{a_{k_n}}{a_{k_n}}^{A_{k_n}^O}
        \notag\\
        &\qquad =
        W_{(k_1,\ldots,k_n)}
        \otimes \id^{A_{k_n}^O}.
        \label{eq:qccc-recursion}
    \end{align}
    The first equality follows because every compatible pair
    $(a_{k_1},...,a_{k_{n-1}},a_{k_n})$ occurs exactly once in the sum, namely in the term
    corresponding to the next party selected by the protocol. The second
    equality follows from the completeness of the classical output basis.
    This is precisely the recursive QC-CC constraint
    in Eq.~\eqref{eq:QCCCchara2}.

    For a complete branch $(k_1,\ldots,k_N)$, take the final system $F$ to be
    one-dimensional and define
    \begin{align}
         W_{(k_1,\ldots,k_N,F)}
        :=
        W_{(k_1,\ldots,k_N)}
        \otimes \id^{A_{k_N}^O}
        \otimes \ketbra{0}{0}^{F}.
    \end{align}
       
    It then follows that Eq.~\eqref{eq:QCCCcharac1} is satisfied.

    Finally, every output string $\bm a$ follows exactly one complete branch
    of the protocol. Expanding the identity on the output of the last party, $\id^{A_{k_N}^O}
        =
        \sum_{a_{k_N}}
        \ketbra{a_{k_N}}{a_{k_N}}^{A_{k_N}^O}$,
    and summing over all complete branches therefore gives
    \begin{align}
        \sum_{(k_1,\ldots,k_N)}
        W_{(k_1,\ldots,k_N,F)}
        &=
        W_{\bm w}^{\bm A}\otimes\ketbra{0}{0}^{F},
    \end{align}
    which is
    Eq.~\eqref{eq:QCCCchara0}.

    All the operators constructed above are positive semidefinite and satisfy
    Eqs.~\eqref{eq:QCCCchara0}--\eqref{eq:QCCCchara3}. Hence
    $W_{\bm w}^{\bm A}$ is a QC-CC process matrix.

    \end{proof}   
\end{theorem}

\begin{corollary}\label{cor:noQCNICC}
    Every classical deterministic QC-CC process that is dynamical, i.e. not QC-FO, has influenceable classical control of causal order.
\end{corollary}

\begin{proof}
    By Theorem~\ref{th:QCCC}, every process matrix of a maximal FR process function is a QC-CC.
    Let us define two sub-classes of QC-CC introduced in Ref.~\cite{mothe2025}. QC-convFO is the convex hull of fixed-order circuits, $W=\sum_{\pi}q_\pi W_\pi,$ with $    q_\pi\geq0,$ $
    \sum_\pi q_\pi=1,$
where each $W_\pi$ is compatible with the fixed order $\pi$. QC-NICC is the
class of QC-CC processes admitting a realization in which the classical
control of the causal order cannot be influenced by the parties' CPTP
operations. In general,
\begin{align}
      \mathrm{QC\text{-}FO}
    \subseteq
    \mathrm{QC\text{-}convFO}
    \subseteq
    \mathrm{QC\text{-}NICC}
    \subseteq
    \mathrm{QC\text{-}CC}.
\end{align}
    It therefore suffices to show that any process belonging to QC-NICC is in QC-convFO. This is straightforward, as
 for deterministic classical process matrices:
\begin{align}
        \mathrm{QC\text{-}convFO}^{det.cl.}
    =
    \mathrm{QC\text{-}NICC}^{det.cl.}
    =
    \mathrm{QC\text{-}FO}^{det.cl.}
\end{align}

Indeed, a deterministic classical process function assigns a unique input
string $\bm x=\bm w(\bm a)$ to every output string $\bm a$. If it admits a
convex decomposition $ W_{\bm w}=\sum_\lambda q_\lambda W_\lambda,$ $q_\lambda>0$,
then each $W_\lambda$ must produce the same assignment $\bm x=\bm w(\bm a)$ for every $\bm a$; otherwise the mixture would assign
nonzero probability to a different input string and would not be
deterministic. Hence every component with nonzero weight realizes the same
process function. It follows that, in a QC-convFO decomposition, each fixed-order component
with nonzero weight realizes $\bm w$, so $\bm w$ is itself compatible with
that fixed order; $\mathrm{QC\text{-}convFO}^{det.cl.}=\mathrm{QC\text{-}FO}^{det.cl.}$. Likewise, in a QC-NICC realization, the distribution of
orders is independent of the parties' operations, and determinism forces
every order occurring with nonzero probability to realize the same function
$\bm w$. Since each complete branch has a fixed
causal order, $\bm w$ must be compatible with every order occurring with
nonzero probability, and therefore with at least one fixed order. Hence
$\mathrm{QC\text{-}NICC}^{det.cl.}
    \subseteq
    \mathrm{QC\text{-}convFO}^{det.cl.}$.
The converse follows from the general inclusion
$\mathrm{QC\text{-}convFO}\subseteq\mathrm{QC\text{-}NICC}$.
\end{proof}

\section{Agency in process functions\label{sec:agency}}

\begin{figure}[ht]
	\begin{center}
	\includegraphics[width=.45\textwidth]{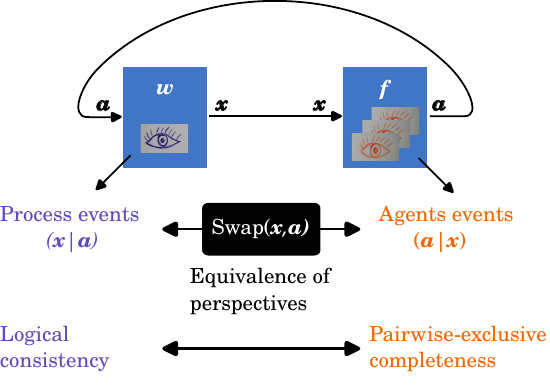}
	\end{center}
	\caption{Interpretation of agency in process functions. 
    Compared to Fig.~\ref{fig:bipartite_pf} and \ref{fig:pf} we employ a coarse-grained view of the retrocausal identity channels and parties: $n$ isolated parties are placed in the right blue box, where each party $k$ receives a classical variable $x_k$ (defining the global input $\bm{x}:=(x_1,...,x_n))$ to which they can apply an arbitrary local operation  $f_k$ (defining a  global operation $\bm{f}:=(f_1,...,f_n)$) and producing an output $a_k$ (defining a  global output $\bm{a}:=(a_1,...,a_n)$). $\bm{a}$ serves as input to the process function $\bm{w}$, which returns   $\bm{x}$ as output. The events from the perspectives of the process function and the local parties, which we denote \emph{process events} and \emph{agents events}, respectively, are related by a simple swap of $\bm{a}$ and $\bm{x}$ (see top arrow). 
    There is a deeper mathematical representation of this connection (see bottom arrow): the process events define a process function that satisfy logical consistency (cf. Def.~\ref{def:mp_pfs})~\cite{baumeler16}, while the agents events 
    that satisfy pairwise-exclusive completeness (Theorem~\ref{thm:characPE})~\cite{dourdent2026makescausalloopconsistent} and define the outcomes of a no-signaling measurement. 
    This work provides a hypergraph representation of this equivalence.}
\label{fig:agency}
\end{figure}

It has as  recently been highlighted by Grinbaum \cite{grinbaum2026} that bridging  indefinite causal order in the operational approach with dynamical theories in spacetime requires to define  agents (or observers) within the theory and in a perspectival way. 
In particular, the global process is treated as prior, while agents are ``admissible groupings of input and output data in a global Hilbert space'' (``Agents from bits''). The original data is unlabeled \textendash\ it is only by dividing the data into inputs and outputs and grouping it  by assigning  labels (e.g., ``Alice'' or ``Bob'') that observers arise.  Redefining these groupings may even eliminate non-causality. In particular, agency is treated as relative to the operational perspectives that result from the definition of agency. 
The grouping, however is not arbitrary: the process needs to satisfy logical consistency that rules out paradoxes, therefore restricting the possible groupings.  This philosophical stance on agency is referred to as \textit{operational eternalism}, in analogy to the eternalist, ``block-universe'' view,  in which all spacetime events are treated on the same footing.

Inspired by this notion of agency, we inspect the process function scenario. 
In particular, we can identify two types of events.  As depicted in Fig.~\ref{fig:agency}$,  
\bm{a}$ and $\bm{x}$ can play different roles: from the perspective of the process, $\bm{a}$ are inputs and $\bm{x}$ outputs, defining a set of \textit{process events} $(\bm{x}|\bm{a})$ with $\bm{x}=\bm{w}(\bm{a})$. 
From the perspective of the parties, $\bm{x}$ are inputs (or rather, measurement settings) and $\bm{a}$ are outputs (or rather, measurement outcomes), defining  \textit{agents events} $(\bm{a}|\bm{x})$. 

Thus process and agents events are ultimately the same events, but they are structured differently: the former globally, from the perspective of the process, the latter locally, from that of a single agent.
On the one hand, process events are uniquely determined by $\bm{a}$ alone, i.e., one can identify them in a coarse-grained, setting-independent manner. On the other hand, identifying the local events $(a_k|x_k)$ of an agent requires both outcomes and settings, i.e., one can identify local events only in a fine-grained, setting-dependent manner. One may then associate agency with respect to these two different perspectives:  process events correspond to a superobserver that is \emph{trivial} in the sense that its outputs $\bm{x}$ can be identified from $\bm{a}$ alone. 
 This global perspective is described as ``observer-indefinite'' in Ref.~\cite{grinbaum2026}. 
The \emph{trivial} events of this global superobserver are themselves the coarse-graining  of the local events that can be associated with local agents.

The equivalence between these two perspectives, related by a simple swap operation (Fig.~\ref{fig:agency}), is grounded by the operational equivalence of their validity: process events define a  function that is logically consistent, while agents events are complete and satisfy pairwise exclusivity.
Our work provides a hypergraph representation of this equivalence.

\end{document}